\documentclass[%
 reprint,
 amsmath,amssymb,
 aps,
 prd,
]{revtex4-2}
\usepackage{url}
\usepackage{footnote}
\usepackage{enumitem}
\usepackage{graphicx}
\usepackage{dcolumn}
\usepackage{bm}
\usepackage{xcolor}
\usepackage{hyperref}
\usepackage{ulem}

\begin{document}

\preprint{APS/123-QED}

\title{N-body Simulations of cosmologies with Light Massive Relics}

\author{Vikhyat Sharma}
\email{vikhyat.sharma@students.iiserpune.ac.in}
\author{Arka Banerjee}
\affiliation{%
 Department of Physics, Indian Institute of Science Education and Research, Homi Bhabha Road, Pashan, Pune 411008, India
}%

\begin{abstract}
\section*{Abstract}
The presence of additional relativistic particles at the time of recombination can be inferred through their contribution to $\Delta N_{\rm eff}$. If these species have a finite but low mass (Light Massive Relics - LiMRs), they act as a hot subcomponent of dark matter and impact late-time structure formation. Understanding these effects will be crucial to pin down the underlying particle physics properties of any future $\Delta N_{\rm eff}$ detection. While their impact has been well-studied on linear scales, this work develops the framework for and presents results from the first set of cosmological N-body simulations that can track the effects of LiMRs, as a function of their mass and temperature, down to fully nonlinear scales. Importantly, our simulations model the impact of both the massive Standard Model neutrinos and LiMRs, which will be crucial in disentangling possible degeneracies. We systematically explore the effects of LiMR properties such as mass, temperature, and initial distribution, on various cosmological observables, including the total matter power spectrum, Halo Mass Functions (HMF), Mass-Concentration relation, radial halo profiles, and weak lensing signals around massive clusters. The framework and simulations developed here will enable detailed follow-up of the rich phenomenology of LiMR cosmologies.

\begin{description}
\item[Keywords]
N-body simulations, Large Scale structures, LiMRs, Structure formation

\end{description}
\end{abstract}

\maketitle


\section{Introduction}
Roughly 85\% of the total matter budget in the universe is ``dark matter" (DM), but its fundamental nature has yet to be understood. Observationally, there is evidence for the majority of dark matter to be made up of cold collisionless particles --- the so-called CDM paradigm\,\cite{Diemand:2011:1936-6612:297, Clowe_2006, 2016, doi:10.1073/pnas.1308716112, Blumenthal1984, Feldmann_2014, 10.1046/j.1365-8711.2003.06207.x, 1970ApJ...159..379R, 10.1093/mnras/stx2630, BERTONE2005279, Nadler_2021}. Nonetheless, the presence of additional subdominant components has not been excluded. In particular, the existence of light relic particles—relativistic at the time of their decoupling in the early universe—in addition to the Standard Model (SM) neutrinos is well motivated by both cosmological considerations and particle physics experiments\,\cite{green2019messengersearlyuniversecosmic, alexander2016darksectors2016workshop, essig2013darksectorsnewlight, archidiacono2013cosmicdarkradiationneutrinos, wallisch_2018}. Any light massive relic (LiMR) species that decoupled in the hot, dense environment of the early universe would contribute to the effective number of relativistic degrees of freedom, appearing as a change in $\Delta N_{\text{eff}}$ at recombination\,\cite{10.48550/arxiv.2101.07810, PhysRevLett.129.021302}. Upcoming CMB experiments are expected to place stringent constraints on $\Delta N_{\rm eff}$\,\cite{abazajian2016cmbs4sciencebookedition, dvorkin2022darkmatterphysicscmbs4}. This makes it important to systematically investigate the impact of both the microscopic properties (such as particle mass and thermal distribution) and the macroscopic properties (such as energy density and free-streaming scale) of LiMRs on cosmological observables across different epochs of the universe. In contrast to massless relics, relics with non-zero masses (LiMRs) behave differently at late times: owing to their finite mass, they contribute a subdominant fraction of the total matter density and thereby influence the growth of cosmic structures. Their presence leaves characteristic signatures in large-scale structure formation, which can be probed through observables such as the total matter power spectrum and the halo mass function. Hence, the defining feature of these LiMRs is that they act like ``dark radiation"\,\cite{PhysRevD.86.043509} in the early universe\,\cite{Bleau_2024} and become non-relativistic at later times, contributing to the dark matter component of the universe. The existence of LiMR particles results in the suppression of the total matter power spectrum at scales smaller than the largest comoving free-streaming scale in the entire history of the universe's evolution. On small scales, these particles fall into the DM potential wells created by non-linear structures such as halos(predominantly sourced by the CDM component)\,\cite{PhysRevD.92.063505}.

While SM neutrinos provide the minimal example of LiMRs, a variety of Beyond-Standard-Model(BSM) particles—such as sterile neutrinos\,\cite{PhysRevD.79.045026, Hagstotz_2021, Aguilar_Arevalo_2013}, axions\,\cite{Marsh_2016}, or other dark-sector species\,\cite{DePorzio_2021, dvorkin2022physicslightrelics}—can also behave as light massive relics. In the minimum-temperature scenario, their contribution to $\Delta N_{\text{eff}}$ values asymptotes to 0.027 for a massless scalar (spin-0), 0.047 for a Weyl fermion (spin-1/2), and 0.054 for a vector boson (spin-1), respectively. Previous studies have investigated the impact of LiMRs primarily on linear scales, using either cosmological survey data\,\cite{PhysRevD.105.095029, banerjee2025forecastingconstraintsnonthermallight, abazajian2012lightsterileneutrinoswhite, annurev:/content/journals/10.1146/annurev.nucl.010909.083654, Baumann_2018, Banerjee_2018_, celik2025mixeddarkmattergalaxy, verdiani2025effectivefieldtheorylarge} or N-body simulations\,\cite{10.1093/mnras/stac2128, Brandbyge_2017}.  However, most existing N-body simulations that include LiMR particles neglect the effects of SM neutrinos, while those that do include both SM neutrinos and LiMRs often fail to capture the full non-linear evolution of all three species. In this work, we develop a methodology to perform N-body simulations that simultaneously evolve CDM, SM neutrinos, and LiMRs, thereby accounting for their complete non-linear dynamics.

N-body simulations require accurate calculations of the matter power spectrum, transfer functions, and growth rates at the initial redshift to establish reliable initial conditions. The Boltzmann solver \texttt{CLASS}\,\cite{DiegoBlas_2011} is widely used for generating these initial conditions in cosmological simulations; however, its direct application to massive neutrino cosmologies presents limitations(for details see\,\cite{10.1093/mnras/stw3340}). To incorporate LiMRs into N-body simulations, it is natural to build upon the state-of-the-art techniques developed for massive neutrinos. Over the past decade, a variety of methods have been proposed to include massive neutrinos in N-body simulations\,\cite{Banerjee_2016, Bayer_2021, _2023, Bird_2012, Brandbyge_2008, Banerjee_2018, Villaescusa_Navarro_2014, Emberson_2017, Adamek_2017, Brandbyge_2009, Liu_2018, Dakin_2019}. Two primary challenges arise in this context: first, the inclusion of massive neutrinos introduces scale-dependent growth rates, necessitating higher accuracy in their calculation; second, at high initial redshifts (e.g., z = 99), a significant fraction of relativistic neutrino particles persists, which are neglected when \texttt{CLASS} is used directly for massive neutrino cases. This shortcoming has been addressed by the development of the \texttt{RePS} code\,\cite{10.1093/mnras/stw3340}, which enables accurate initial condition generation in massive neutrino cosmologies. Prominent simulation suites, such as the Quijote simulations\,\cite{Villaescusa-Navarro_2020}, the Euclid emulator\,\cite{10.1093/mnras/stab1366}, and the Bacco emulator\,\cite{10.1093/mnras/stab2018}, employ \texttt{RePS} for setting up initial conditions when neutrinos are included. Since LiMRs share many cosmological features with massive neutrinos—such as free-streaming, scale-dependent suppression of clustering, and a relativistic-to-nonrelativistic transition—the techniques developed for neutrinos provide a natural framework for extending N-body methods to LiMRs. Building on the capabilities of \texttt{RePS}, we have developed a methodology to compute accurate initial conditions for N-body simulations that simultaneously evolve CDM + baryons, SM neutrinos, and LiMRs. Developing such a methodology is essential for two main reasons. First, it allows us to quantify the unique signatures of LiMRs on cosmological observables, which is crucial for breaking degeneracies between LiMRs and SM neutrinos. Second, it provides the accuracy needed to model the small-scale effects of LiMRs on the matter power spectrum, halo properties, and weak-lensing observables at late times. By combining precise N-body predictions with both early- and late-time cosmological probes, we can extract substantially more information about LiMRs and improve our ability to constrain their properties from upcoming surveys.

Section II describes the methodology developed to extend the multi-fluid formalism of \texttt{RePS}, originally designed for massive neutrinos, to include LiMR particles for the accurate generation of initial conditions (ICs). Modifications implemented to enable N-body simulations with CDM, SM neutrinos, and LiMR particles are described in detail. In this work, we investigate thermal LiMRs with masses of a few eV. Such particles exhibit rich phenomenology at the intersection of particle physics and cosmology. They are neither heavy enough to cluster like CDM nor light enough to remain relativistic until the present epoch, thereby influencing large-scale structure formation in distinctive ways. Since LiMRs can, in principle, follow different distribution functions(based on spin of the particles), we model them first as Dirac fermions(spin-1/2) and subsequently as vector bosons(spin-1), demonstrating the flexibility of our framework in incorporating different particle statistics and distribution functions. In the massless limit, and under the minimum-temperature scenario, these yield  $\Delta N_{\text{eff}}$ values of 0.094 and 0.054, respectively. Section III presents the results of our simulations, focusing on a range of cosmological observables: the total matter power spectrum, the halo mass function (HMF), the mass-concentration relation of dark matter halos, the three-dimensional halo density profiles, and weak-lensing statistics. Section IV concludes with a summary of our findings and outlines possible directions for future research.

\section{Methodology}
We adopt the following terminology throughout this work. The total matter density parameter is denoted by $\Omega_m$, while $\Omega_{cb}$, $\Omega_\nu$, and $\Omega_l$ represent the density parameters of CDM+baryons, SM neutrinos, and LiMRs, respectively. A cosmology in which the matter content consists solely of CDM and baryons is referred to as the 1-fluid case. When SM neutrinos are included alongside CDM+baryons, the system is treated as two distinct fluids, which we denote as the 2-fluid case. Finally, when LiMRs are also incorporated, the matter content is described by three distinct fluids that evolve simultaneously, and we refer to this as the 3-fluid case.

\begin{itemize}
  \item 1-fluid case $\longrightarrow \Omega_m=\Omega_{cb}$
  \item 2-fluid case $\longrightarrow \Omega_m=\Omega_{cb}+\Omega_\nu$ 
  \item 3-fluid case $\longrightarrow \Omega_m=\Omega_{cb}+\Omega_\nu+\Omega_l$
\end{itemize}

For the 1-fluid case, the \texttt{RePS} code takes the output of \texttt{CLASS} and modifies the results using a backscaling method, ensuring that subsequent simulations with the \texttt{Gadget-3} code yield accurate results. When addressing the massive neutrino case, the code not only resolves the issue of incorrect background expansion but also simultaneously calculates the correct scale-dependent growth rates for neutrinos. Extending the formalism in the \texttt{RePS} code, we have developed a backscaling technique to calculate accurate initial conditions for 3-fluid simulations, which include CDM + baryons, SM neutrinos, and LiMRs. In the following section, we detail the model employed for backscaling to generate precise initial conditions by modifying the \texttt{CLASS} output at redshift 99, specifically for the 3-fluid case. Furthermore, we present the equations required for setting up the free streaming scales ($k_{fs}$) for SM neutrinos and LiMRs, which are required to calculate the accurate initial conditions at z=99. We also present the equations required to compute density parameters($\Omega_l$), and particle velocities for the thermal LiMR particles as they are required to set up the initial conditions for N-body simulations, as discussed in section III.

\subsection{Formalism}
The fluid equations are formulated using perturbation theory, retaining terms up to linear order. For cold dark matter (CDM), the pressure term is absent, and gravitational forces solely drive the growth of structures. After recombination, baryonic pressure becomes negligible on the large scales relevant for our study, and by neglecting additional non-gravitational interactions, the evolution of baryonic density fluctuations effectively follows that of CDM. Consequently, throughout this work, we treat CDM and baryons as a single effective fluid. The evolution of their density perturbations is then governed by the continuity and Euler equations. For CDM, these equations are expressed as follows\,\cite{knobel2012introduction}:

\begin{equation}
    \frac{\partial \delta_{cb}}{\partial t}+\frac{\theta_{c b}}{a}=0 \quad,
    \label{eq:a}
\end{equation}

\begin{equation}
    \frac{\partial \theta_{cb}}{\partial t}+H \theta_{cb}=-\frac{1}{a} \nabla^2 \phi \quad .
\end{equation}
where, $\delta=(\rho - \bar{\rho})/\bar{\rho}\quad, \theta=\vec{\nabla} \cdot \vec{v}$. SM neutrinos retain significant thermal velocities from their relativistic decoupling. As a result, they do not cluster efficiently on small scales, since they can stream out of the potential wells generated by CDM and baryons. In the fluid approximation, this free-streaming manifests as an effective pressure contribution in the Euler equation, corresponding to a nonzero sound speed for neutrinos($c_{s,\nu}$)\cite{PhysRevD.82.089901}. The fluid equations for SM neutrinos are:

\begin{equation}
 \frac{\partial \delta_\nu}{\partial t}+\frac{\theta_\nu}{a}=0 \quad,
\end{equation}

\begin{equation}
    \frac{\partial \theta_\nu}{\partial t}+H \theta_\nu=-\frac{1}{a} \nabla^2 \phi-\frac{c_{s,\nu}^2}{a} \nabla^2 \delta_\nu \quad .
\end{equation}
In a manner similar to neutrinos, LiMR particles also have a non-zero pressure term at the initial redshift due to their thermal velocities. This allows us to define a corresponding sound speed, denoted as $c_{s,l}$.

\begin{equation}
    \frac{\partial \delta_l}{\partial t}+\frac{\theta_l}{a}=0 \quad,
\end{equation}

\begin{equation}
    \frac{\partial \theta_l}{\partial t}+H \theta_l=-\frac{1}{a} \nabla^2 \phi-\frac{c_{s, l}^2}{a} \nabla^2 \delta_l \quad.
\end{equation}
We then use the Poisson equation(which relates gravitational potential to total matter perturbations) to close the above set of equations for 3 different types of particles. Through the Poisson equation, the presence of LiMR particles as well as SM neutrinos affects the clustering of CDM particles because it depends on the total matter overdensity($\delta_m$).

\begin{equation}
   \nabla^2 \phi=\frac{3}{2} \Omega_m \delta_m a^2 H^2    \quad ,
   \label{eq:b}
\end{equation}
where,
\begin{equation}
   \delta_m=f_{cb} \delta_{cb}+f_\nu \delta_\nu+f_l \delta_l \quad .
\end{equation}
Here, $f_{i}$ denotes the mass fractions of each type of matter particle. Light relic species (with mass $\sim$few eV) as well as massive neutrino particles cannot cluster efficiently on scales larger than their free streaming scales\,\cite{10.1088/1475-7516/2017/07/033, 10.1103/physrevd.77.023528} and freely stream out of potential wells created by CDM+baryons. The sound speeds ($c_{s,i}$) corresponding to the SM neutrinos and LiMR particles can be related to their respective free streaming scales ($k_{fs,i}$) as follows\,\cite{Blas_2014}:

\begin{equation}
   c_{s, \nu}^2=\frac{3}{2} \Omega_m(a) \frac{a^2 H^2}{\left(k_{fs,\nu}\right)^2}  \quad ,
\end{equation}

\begin{equation}
   c_{s, l}^2=\frac{3}{2} \Omega_m(a) \frac{a^2 H^2}{\left(k_{fs,l}\right)^2} \quad .
   \label{eq:cs_l}
\end{equation}

We solve the above set of equations\,~(\ref{eq:a})--(\ref{eq:b}) in Fourier space using the following normalizations $\quad \delta_i(a, \vec{k})= D_i(a, \vec{k}) N(\vec k)$ and $\quad \theta_i(a, \vec{k})= a H(a) X_i(a, \vec{k}) N(\vec k) \ $ where $i$ represents different types of particle species. Solving these equations, we obtain

\begin{equation}
    \frac{\partial D_{cb}}{\partial \ln a}=-X_{cb}  \quad ,
    \label{eq:c}
\end{equation}

\begin{equation}
    \frac{\partial X_{cb}}{\partial \ln a}=A X_{cb}+B\left(f_{cb} D_{cb}+f_\nu D_\nu+f_l D_l\right)  \quad ,
\end{equation}

\begin{equation}
    \frac{\partial D_\nu}{\partial \ln a}=-X_\nu  \quad ,
\end{equation}

\begin{multline}
    \frac{\partial X_\nu}{\partial \ln a}=A X_\nu+ \\
    B\left[f_{cb} D_{cb} +\left(f_\nu-\frac{k^2}{\left(k_{fs,\nu}\right)^2}\right) D_\nu +f_l D_l \right]  \quad ,
\end{multline}

\begin{equation}
    \frac{\partial D_l}{\partial \ln a}=-X_l  \quad ,
\end{equation}

\begin{multline}
    \frac{\partial X_l}{\partial \ln a}=A X_l+ \\
    B\left[f_{c b} D_{c b}+f_\nu D_\nu+\left(f_l-\frac{k^2}{\left(k_{fs,l}\right)^2}\right) D_l\right]  \quad .
\end{multline}
where
\begin{equation}
    A=-\left(2+\frac{1}{H^2} \frac{dH}{dt}\right)  \quad ,
\end{equation}
\begin{equation}
    B=-\frac{3}{2} \Omega_m(a) \quad .
    \label{eq:d}
\end{equation}

We now turn to the calculation of the LiMR quantities that enter the equations above, namely the free-streaming scale, energy density, and velocity dispersion. These quantities are derived from the fundamental particle properties of LiMRs—such as their mass, temperature, and distribution function—as outlined below.

\subsubsection*{Sound speed and Free streaming scales}
The sound speed of a light relic species can be related to its velocity dispersion(as in \,\cite{PhysRevD.82.089901}) using the following relation:-

\begin{equation}
    c_s^2 = \frac{5}{9} \ \sigma_v^2 \quad .
\end{equation}
Since fermions and bosons have different velocity distribution functions, their velocity dispersion differs by a factor of $\sqrt{4/5}$ and can be written as:-

\begin{equation}
    \sigma_v^{\rm boson} = \sqrt{\frac{4}{5}} \ \sigma_v^{\rm fermion} \quad .
\end{equation}
Using the velocity dispersion formula for relic species that follows the Fermi-Dirac distribution, or the Bose-Einstein distribution, the corresponding sound speed can be calculated as:

\begin{equation}
    c_s^{\rm fermion} \approx 68.836 \cdot (1+z) \cdot \left( \frac{T_{i,0}}{1 K} \right) \cdot \left( \frac{1eV}{m_i} \right) \quad ,
\end{equation}

\begin{equation}
    c_s^{\rm boson} \approx 61.5688 \cdot (1+z) \cdot \left( \frac{T_{i,0}}{1 K} \right) \cdot \left( \frac{1eV}{m_i} \right) \quad .
\end{equation}
Here $T_{i,0}$ is the temperature of the relic species today in K and $m_i$ is its mass in eV. Corresponding to this sound speed, the free streaming scale can be calculated using equation\,\eqref{eq:cs_l}.

\begin{equation}
    k_{fs}^{\rm fermion} = \sqrt{\frac{3}{2} \Omega_m(a)} \ \cdot \frac{a^2H}{68.836} \cdot \left( \frac{1 K}{T_{i,0}} \right) \cdot \left( \frac{m_i}{1eV} \right) \quad ,
\end{equation}

\begin{equation}
    k_{fs}^{\rm boson} = \sqrt{\frac{3}{2} \Omega_m(a)} \ \cdot \frac{a^2H}{61.5688} \cdot \left( \frac{1 K}{T_{i,0}} \right) \cdot \left( \frac{m_i}{1eV} \right) \quad .
\end{equation}

\begin{figure*}
  \centering
  \includegraphics[width=\linewidth]{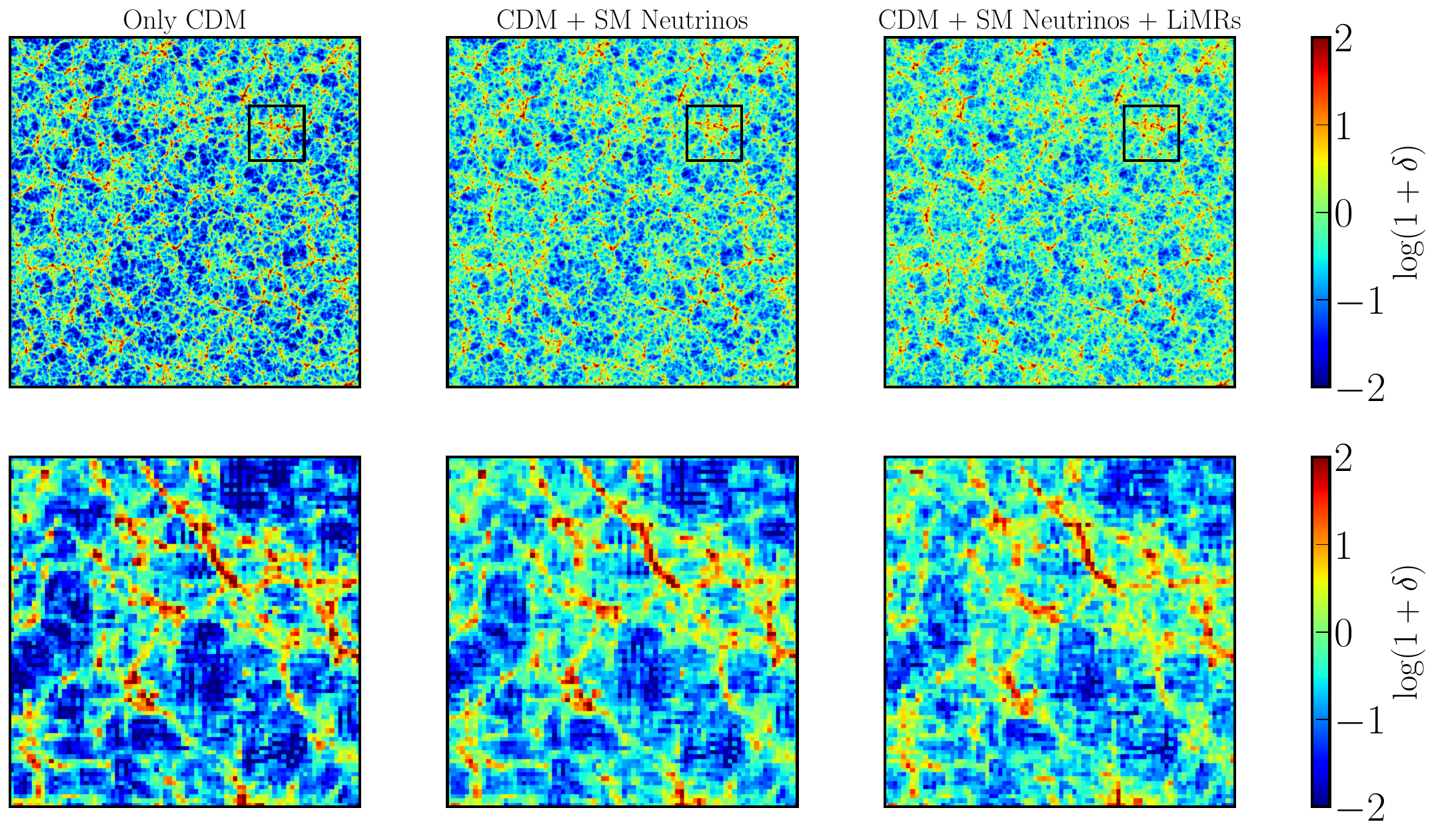}  
  \caption{This is a 2D projected density field generated by taking the mean along the z-axis for a slice of width 10 $h^{-1}\,\mathrm{Mpc}$ for 3 different simulations. The top leftmost panel shows the density field generated for a Only CDM(1-fluid) with total $\Omega_m = \Omega_c = 0.3175$. In the middle panel is the density field of a simulation with CDM and SM neutrinos(2-fluid) with total $\Omega_m = (\Omega_c + \Omega_{\nu}) = 0.3175$ for $\Sigma m_{\nu} = 0.60$ eV. Similarly, the top right-most panel corresponds to a simulation having CDM, SM neutrinos, and LiMRs(3-fluid) with total $\Omega_m = (\Omega_c + \Omega_{\nu} + \Omega_l) = 0.3175$ for $\Sigma m_{\nu} = 0.60$ eV and a single species of LiMR particle having mass 5 eV. The lower 3 plots are zoomed-in plots for the boxes marked in the corresponding density plots above them.}
  \label{fig:den_tot}
\end{figure*}

\subsubsection*{Density parameters of LiMR particles}
Matter density of a thermal LiMRs, that decoupled while being relativistic, can be calculated as follows:-
\begin{equation}
\Omega_i^{\rm fermions} = \left( \frac{m_\nu}{93.14 h^2} \right) \cdot \left( \frac{m_i}{m_\nu} \right) \cdot \left( \frac{g_i}{g_\nu} \right) \cdot \left( \frac{T_{i,0}}{T_{\nu,0}} \right)^3 \quad ,
\label{eq:omega_f}
\end{equation}

\begin{equation}
\Omega_i^{\rm bosons} = \frac{4}{3}\times\left( \frac{m_\nu}{93.14 h^2} \right) \cdot \left( \frac{m_i}{m_\nu} \right) \cdot \left( \frac{g_i}{g_\nu} \right) \cdot \left( \frac{T_{i,0}}{T_{\nu,0}} \right)^3 \ .
\end{equation}
where $m_i$ is the mass of a single LiMR species in eV, $T_{i,0}$ is the temperature of the particles today, and $g_i$ is the number of internal degrees of freedom that were in equilibrium while decoupling. This has been written w.r.t. mass of a single neutrino species($m_\nu$) and the neutrino temperature($T_{\nu,0}$) today. The density parameter for bosons has an additional factor of 4/3, which comes due to its different distribution functions in comparison to fermions.

\subsubsection*{Thermal velocities of LiMR particles}
When LiMR particles decouple from the thermal bath in the early universe, they had relativistic velocities, which decrease as the universe expands. Although their thermal velocities remain high even at later epochs in the universe($z\sim99$), leading to reduced clustering as compared to the CDM particles. We calculate the thermal velocities of each LiMR particle in the simulation box using the following equations:- 

\begin{equation}
    V_i^{\rm fermion}=25.792(1+z)\left(\frac{1 eV}{m_i}\right)\left(\frac{T_{i, 0}}{1 K}\right) \times \tilde{x} \ \ \mathrm{km} / \mathrm{s} \quad ,
    \label{eq:thermal_vel_f}
\end{equation}

\begin{equation}
    V_i^{\rm boson}=25.792(1+z)\left(\frac{1 eV}{m_i}\right)\left(\frac{T_{i, 0}}{1 K}\right) \times \tilde{x} \ \ \mathrm{km} / \mathrm{s} \quad .
    \label{eq:thermal_vel_b}
\end{equation}
where $m_i$ and $T_{i,0}$ denote the mass and present-day temperature of the LiMR species, and $\tilde{x} \in [0,1]$ is a uniform random number that is mapped through the cumulative distribution function (CDF) of the relevant thermal distribution. For fermionic LiMRs we use the Fermi–Dirac distribution, while for bosonic LiMRs we adopt the Bose–Einstein distribution.

\begin{figure*}  
  \centering
  \includegraphics[width=\linewidth]{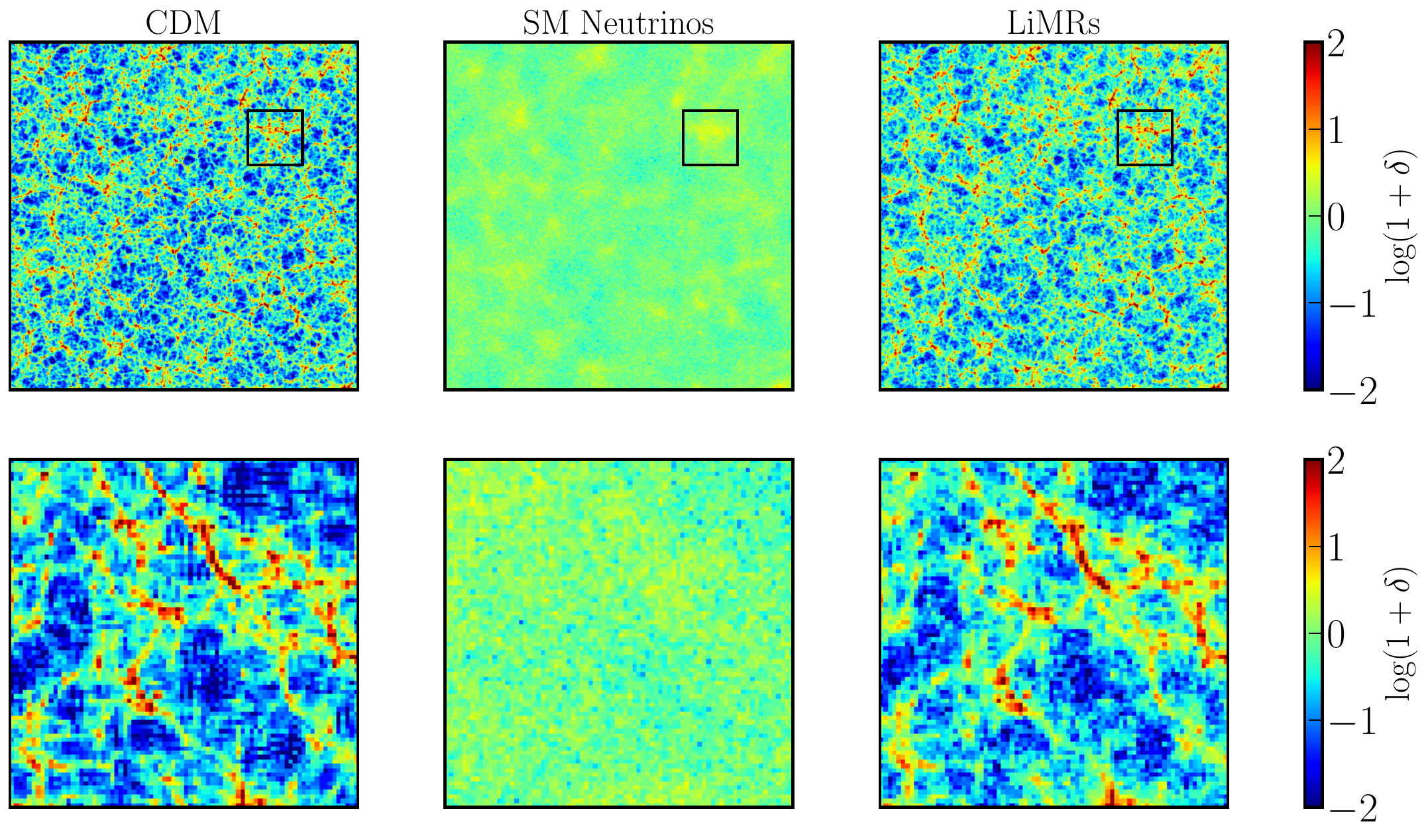}  
  \caption{2D projected density fields obtained by averaging along the z-axis over a slice of width 10 $h^{-1}\,\mathrm{Mpc}$ for the three particle species in the 3-fluid simulation containing CDM, SM neutrinos, and LiMRs. The total matter density parameter is $\Omega_m = \Omega_c + \Omega_\nu + \Omega_l = 0.3175$, with $\Sigma m_\nu = 0.30$ eV and a single LiMR species of mass 1.5 eV. The top row shows the individual projected density fields of CDM (left), SM neutrinos (center), and LiMRs (right), while the bottom row presents zoomed-in views of the regions marked above. LiMRs are seen to cluster less efficiently than CDM but more strongly than SM neutrinos, which show minimal clustering due to their large thermal velocities.}
  \label{fig:den_indi}
\end{figure*}

\subsection{Simulation pipeline}
The methodology outlined in the previous subsection is implemented in a new code, \texttt{Py-RePS}(see \url{https://github.com/vikhyat108/Py-RePS}), which extends the capabilities of \texttt{RePS} to incorporate LiMRs in addition to CDM+baryons and SM neutrinos. \texttt{Py-RePS} computes the modified power spectra, scale-dependent growth rates, and transfer functions for each species at the initial redshift. These outputs are then provided to \texttt{N-GenIC}, which generates the initial conditions (ICs) at z=99. Specifically, \texttt{N-GenIC} employs the Zel’dovich approximation\,\cite{1970AA.....5...84Z} to displace particle positions and assigns peculiar velocities using the scale-dependent growth rates and growth factors obtained from \texttt{Py-RePS}. In this setup, the simulations consistently evolve three particle species: cold dark matter (CDM), Standard Model (SM) neutrinos, and light massive relics (LiMRs). CDM particles are assigned zero thermal velocity. In contrast, SM neutrinos and LiMRs are assigned both thermal and peculiar velocities. The thermal velocities are determined by the expressions\,\eqref{eq:thermal_vel_f} and\,\eqref{eq:thermal_vel_b} in Section II and depend on mass, temperature, and the relevant distribution function. Bosons follow the Bose-Einstein distribution, while fermions follow the Fermi-Dirac distribution. These thermal velocities are added to the peculiar velocities for SM neutrinos and LiMRs, with directions assigned randomly within a sphere. LiMRs generally exhibit lower velocities than SM neutrinos due to their higher masses and lower temperatures. Incorporating thermal velocities for LiMRs significantly influences their clustering properties.

As discussed in Section I, most existing N-body studies of LiMR particles neglect the effects of Standard Model (SM) neutrinos. Although some works have included both massive neutrinos and a hot dark matter (HDM) species, such as sterile neutrinos, none have followed the full nonlinear evolution of a third particle species (LiMR or HDM) in addition to SM neutrinos and cold dark matter (CDM)\,\footnote{A related approach was adopted in\,\cite{pierobon2024tricktreatallsupereasy}, where the non-linear evolution was computed only for slow-moving HDM particles, with fast components treated via linear response\,\cite{Chen_2023}. They further considered a single effective HDM species\,\cite{upadhye2024hotonceneutrinoshot} representing all three SM neutrinos and a light QCD axion ($m_a \lesssim 0.24 \rm{eV
}$). For particles with masses of a few eV (or for colder species), however, this single-species approximation may become inaccurate, as their physical properties can deviate noticeably from those of SM neutrinos.}. A common alternative to the full nonlinear evolution is the grid-based method\,\cite{10.1093/mnras/sts286, Chen_2021}, in which the evolution of the additional species is computed semi-analytically by evaluating gravitational forces from the CDM potential. While computationally efficient, this method relies on the linear response approximation and therefore cannot fully capture nonlinear effects. For particles with masses in the eV range, we examined the impact of explicitly computing short-range forces in N-body simulations. We find that including short-range forces for LiMRs alters the total matter power spectrum by a few percent, even at redshifts as high as z$\sim$99. In contrast, explicitly calculating these forces for SM neutrinos produces negligible differences, owing to their smaller masses and higher thermal velocities. These results indicate that accurate modeling of LiMRs requires short-range force calculations for full nonlinear evolution; hence, the linear response approach is insufficient. To address this, we have modified \texttt{Gadget-3}\,\cite{10.1111/j.1365-2966.2005.09655.x} to self-consistently evolve all three components—CDM, SM neutrinos, and LiMRs—under gravitational interactions. In particular, we incorporate short-range force calculations for LiMRs from the initial redshift (z=99) onward, as these contributions are crucial for capturing their clustering behavior across cosmic time.

Code validation is performed through a consistency check, detailed in the appendix. Two simulations are conducted: one with two particle types, consisting of one species of cold dark matter (CDM) and three species of Standard Model (SM) neutrinos, and another with one species of CDM, two species of SM neutrinos, and the third neutrino designated as a LiMR particle in the \texttt{N-GenIC} and \texttt{Gadget-3} codes. The total energy densities are maintained identically in both simulations, and the temperature and mass of the SM neutrino are assigned to the third LiMR particle. If the modified \texttt{N-GenIC} and \texttt{Gadget-3} codes function correctly, both scenarios should produce equivalent results. Further details are provided in the appendix.

\section{Results}

\begin{figure*}
  \centering
  \includegraphics[width=\linewidth]{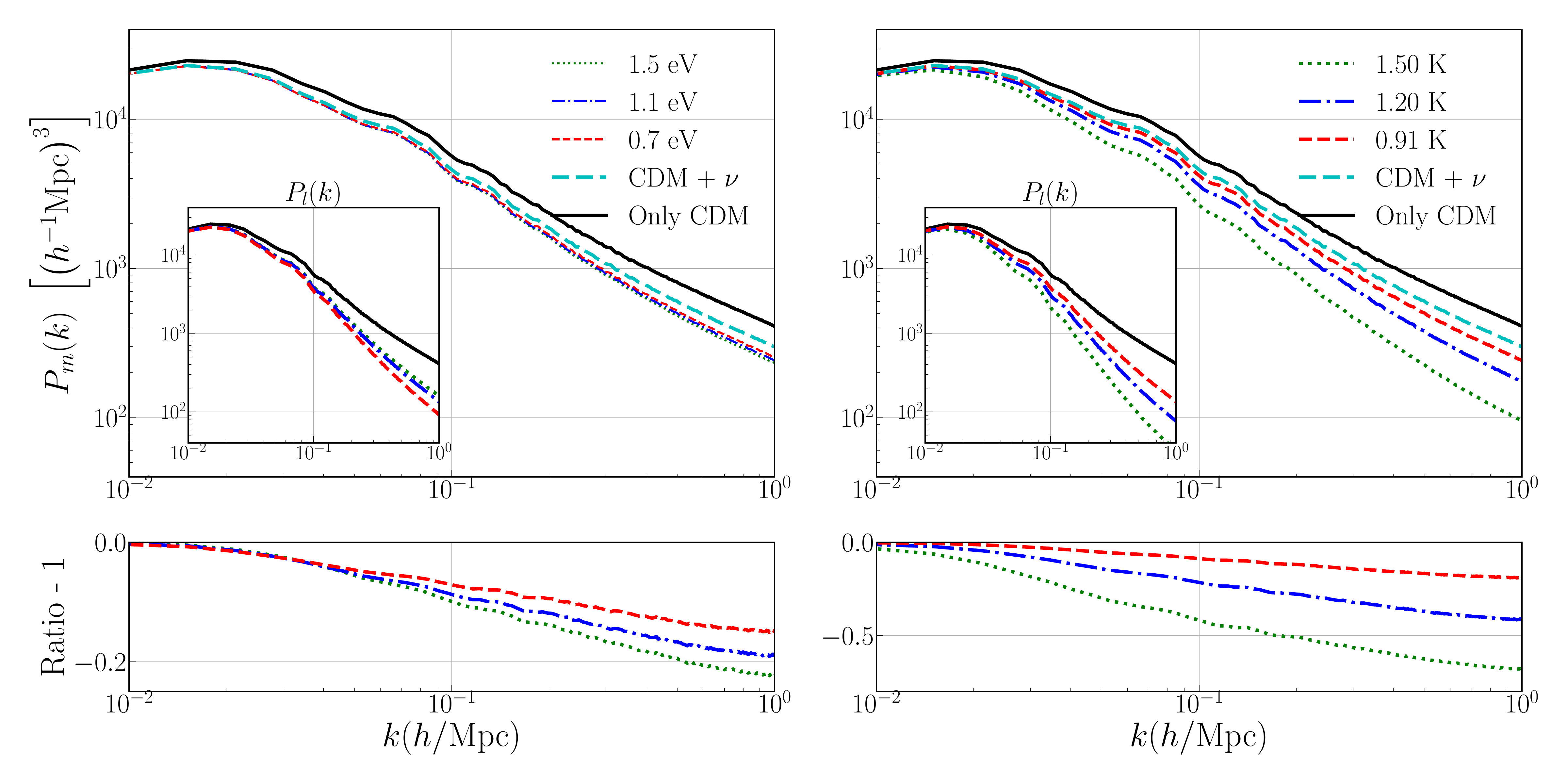}  
  \caption{The top-left panel shows the total matter power spectrum for five different cases: the black(solid) curve corresponds to a 1-fluid model (CDM only), the cyan(long dashed) curve to a 2-fluid model (CDM + SM neutrinos), and the red(dashed), blue(dash-dotted), and green(dotted) curves to 3-fluid models (CDM + SM neutrinos + LiMRs) with LiMR masses of 0.7 eV, 1.1 eV, and 1.5 eV, respectively. The inset displays the LiMR power spectrum($P_l(k)$) for the three LiMR masses, with the black curve showing the CDM power spectrum in the 1-fluid case for comparison. The bottom-left panel compares the total matter power spectrum in the 3-fluid LiMR cosmologies with the total matter power spectrum in the 2-fluid case. On the right, the top panel fixes the LiMR mass at 1.1 eV and varies the LiMR temperature (0.91 K in red (dashed), 1.20 K in blue (dash-dotted), and 1.50 K in green (dotted)), while the bottom panel shows the residuals relative to the 2-fluid case. The corresponding inset highlights the LiMR power spectrum for the three different temperatures. Overall, these results demonstrate that the suppression in the power spectrum becomes stronger with increasing LiMR mass or temperature.}
  \label{fig:pk_f}
\end{figure*}

After modifying the codes and performing the required checks, we run the first cosmological N-body simulations for LiMR cosmology, incorporating SM neutrinos, using our modified pipeline: \texttt{Py-RePS} \textrightarrow \texttt{N-GenIC} \textrightarrow \texttt{Gadget-3}. Figure\,\ref{fig:den_tot} presents density plots that compare the results of CDM (1-fluid), CDM + SM neutrinos (2-fluid), and CDM + SM neutrinos + LiMRs (3-fluid) simulations. As observed in the zoomed-in lower panels, the inclusion of SM neutrinos visibly suppresses large-scale structure formation. Including LiMRs exhibits a similar but even stronger effect on the large-scale structure, further erasing structures at late times. Here, we adopt $\Sigma m_{\nu} = 0.60$ eV and a LiMR mass of 5 eV for illustration. In other analyses, we use different values of $\Sigma m_{\nu}$ and LiMR mass. Figure \ref{fig:den_indi} shows the individual density distributions of the three particle species for a simulation with $\Sigma m_{\nu} = 0.30$ eV and a LiMR mass of 1.5 eV. We choose a lower LiMR mass here because the difference in clustering between heavier LiMR particles and CDM is less pronounced (again, these values are used purely for illustration). Among the three species, CDM exhibits the strongest clustering, SM neutrinos cluster only weakly, and LiMRs—being colder than SM neutrinos but still possessing non-zero thermal velocities (unlike CDM)—display an intermediate degree of clustering.

In the following subsections, we use 3-fluid simulations that have a box size of 1 $h^{-1} \ \mathrm{Gpc}$ with $512^3$ particles. The following cosmological parameters are used in all simulations: $\Omega_m = 0.3175$, $\Omega_b = 0.049$, $\Omega_{\Lambda} = 0.6825$, $h = 0.6711$, $n_s = 0.9624$, and $A_s = 2.13 \times 10^{-9}$. The parameters varied between simulations are $\Omega_{\nu}$, $\Omega_l$, and $\Omega_{c} = \Omega_m - (\Omega_b + \Omega_{\nu} + \Omega_l)$. The values of $\Omega_{\nu}$ and $\Omega_l$ are determined from their respective temperatures and masses using the formulas in Section II. In these simulations, we set the gravitational softening lengths to 1/40 of the mean inter-particle separation for all three types of particles. For all simulations, we fix the sum of neutrino masses to $\Sigma m_{\nu} = 0.30$ eV, with three neutrinos each of mass 0.1 eV and $T_{\nu,0} = 1.95$ K. The LiMR particle is treated as a Dirac fermion, and we conduct five N-body simulations in total. First, the LiMR temperature is fixed at 0.91 K, the minimum temperature a relic species may have today, and three different LiMR masses are considered: 0.7 eV, 1.1 eV, and 1.5 eV. Next, the LiMR mass is fixed at 1.5 eV while varying the temperature to 1.20 K and 1.50 K. This approach allows us to examine the impact of LiMR particle mass and temperature on large-scale structure formation. The chosen ranges for the LiMR mass (0.7–1.5 eV) and temperature (0.91–1.50 K) are well motivated by the parameter space explored in \cite{PhysRevD.105.095029}. Particles in this mass regime occupy an intermediate scale between SM neutrinos and CDM: they are too heavy to remain fully relativistic at late times, yet not massive enough to cluster like CDM on small scales.

When Standard Model (SM) neutrinos and LiMRs are included in the cosmology while keeping the total matter density parameter $\Omega_m$ fixed, the contribution from cold dark matter plus baryons ($\Omega_{cb}$) is reduced. Since CDM is the most efficiently clustering component of matter, a smaller $\Omega_{cb}$ naturally suppresses the overall growth of cosmic structures. This suppression manifests in several large-scale structure observables, such as the total matter power spectrum, the halo mass function\,\cite{10.1093/mnras/stt829}, and related statistics. The presence of additional species such as SM neutrinos or LiMRs further amplifies this suppression. Being relativistic at early times, these species free-stream out of density perturbations and fail to cluster efficiently. Even at late times, when they become non-relativistic, their substantial residual thermal velocities prevent them from clustering on scales larger than their free-streaming lengths. On smaller scales, where they can fall into the gravitational potential wells generated by CDM and baryons, their clustering remains inefficient compared to CDM because of their high velocity dispersion. The suppression of the power spectrum when adding SM neutrinos, while keeping total $\Omega_m$ fixed, is well established in the literature\,\cite{LESGOURGUES2006307, Zeng_2019, https://doi.org/10.1155/2012/608515}. Adding LiMRs produces a similar effect, as these particles are expected to be colder and more massive than neutrinos. Their lower thermal velocities allow them to cluster more than neutrinos, but still less than CDM particles, as discussed in the following subsections. There are two related effects: first, increasing the LiMR mass or temperature raises $\Omega_l$ and reduces $\Omega_{cb}$, leading to less structure formation(we call this the ``effect of reduced CDM fraction"). Second, the change in LiMR clustering due to change in their mass or temperature, we call this the ``effect of LiMR's clustering". In the following subsections, we analyze key observables, including the total matter power spectrum ($P(k)$), halo mass functions (HMF), the mass-concentration relation of halos, 3D density profiles of DM halos, and weak lensing, as discussed in detail in the following subsections. Results from the 3-fluid simulations are compared with the 1-fluid (CDM-only) and 2-fluid (CDM + SM neutrinos) cases to isolate the impact of introducing LiMRs as a third particle species.

\begin{figure*}
  \centering
  \includegraphics[width=\linewidth]{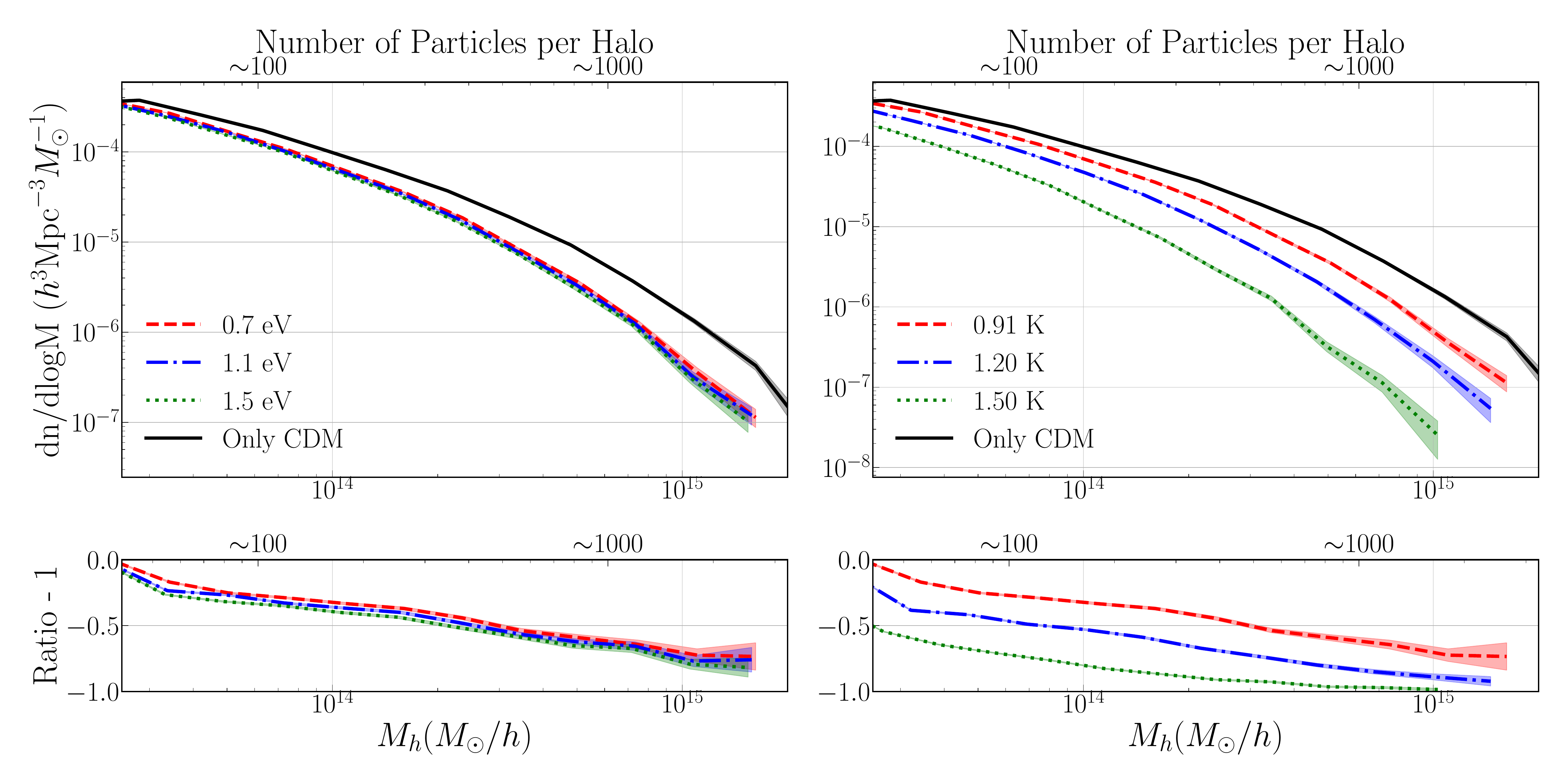}  
  \caption{Halo mass function for the 3-fluid model. The black solid line corresponds to a 1-fluid cosmology (CDM only, without SM neutrinos or LiMRs) and is compared to the 3-fluid case (CDM + SM neutrinos + LiMRs). In the left column, the red dashed, blue dash–dotted, and green dotted lines represent LiMR masses of 0.7 eV, 1.1 eV, and 1.5 eV, respectively (assuming Dirac fermions). In the right column, the LiMR mass is fixed at 1.1 eV, while the temperature is varied: 0.91 K (red dashed), 1.20 K (blue dash–dotted), and 1.50 K (green dotted). The shaded regions indicate the $1\sigma$ uncertainties. The bottom row shows the ratio of the 3-fluid halo mass function to that of the 1-fluid case.}
  \label{fig:hmf_f}
  
  \vspace{0.1em} 
  
  \centering
  \includegraphics[width=\linewidth]{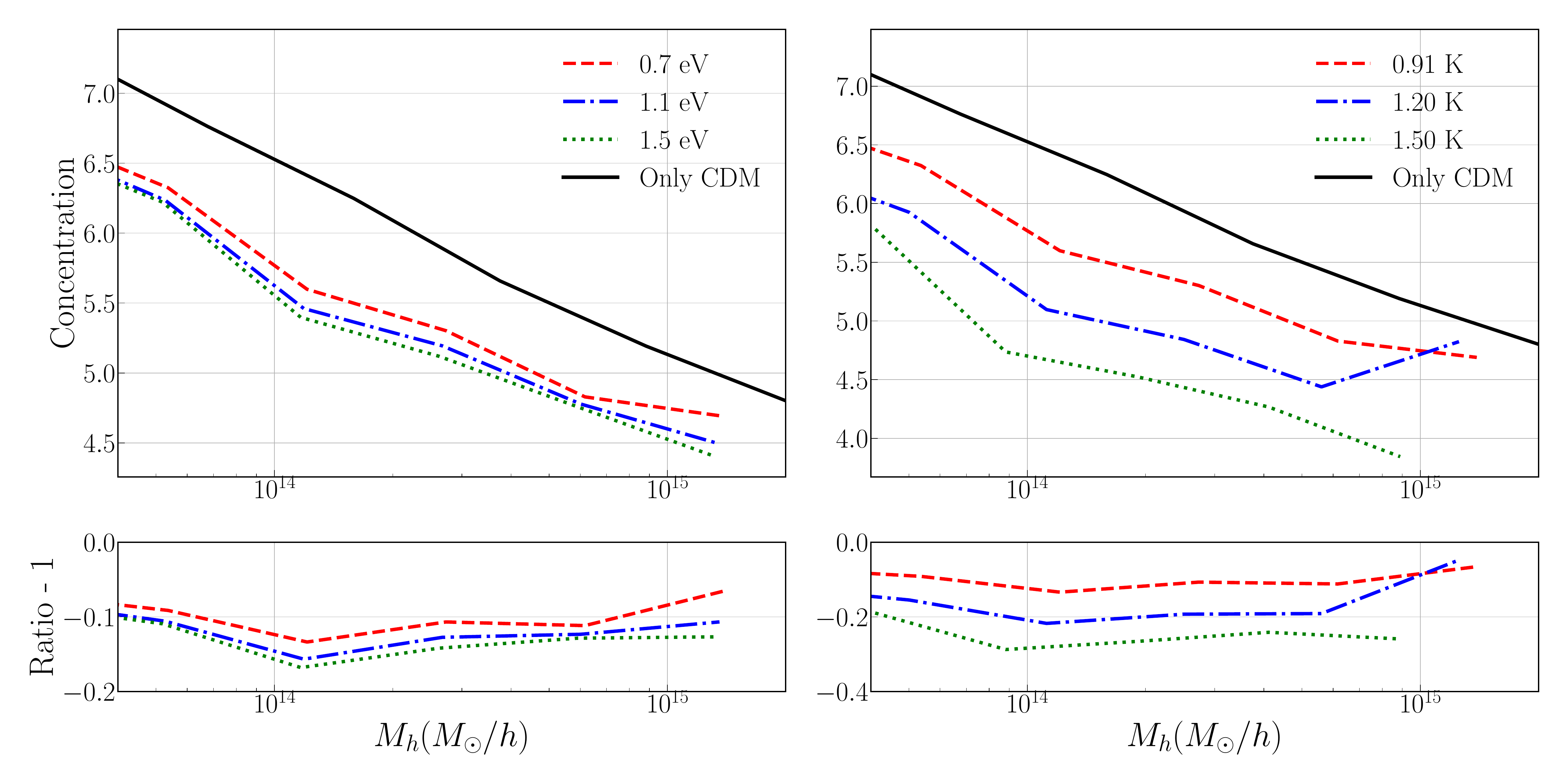}  
  \caption{Mass–concentration relation for the 3-fluid model. The black solid line corresponds to a cosmology with only CDM. In the top-left panel, the red dashed, blue dash-dotted, and green dotted lines represent the 3-fluid case (CDM + SM neutrinos + LiMRs) with a fixed LiMR temperature of 0.91 K and LiMR masses of 0.7 eV, 1.1 eV, and 1.5 eV, respectively. The bottom-left panel shows the relative suppression of the mass–concentration relation in the 3-fluid model with respect to the CDM-only case. The results indicate that the suppression becomes stronger with increasing LiMR mass. In the top-right panel, the red dashed, blue dash-dotted, and green dotted lines correspond to the 3-fluid model with a fixed LiMR mass of 1.1 eV and temperatures of 0.91 K, 1.20 K, and 1.50 K, respectively. The bottom-right panel shows the corresponding relative suppression with respect to the CDM-only (1-fluid) case.}
  \label{fig:conc_f}
\end{figure*}

\begin{figure*}
  \includegraphics[width=\linewidth]{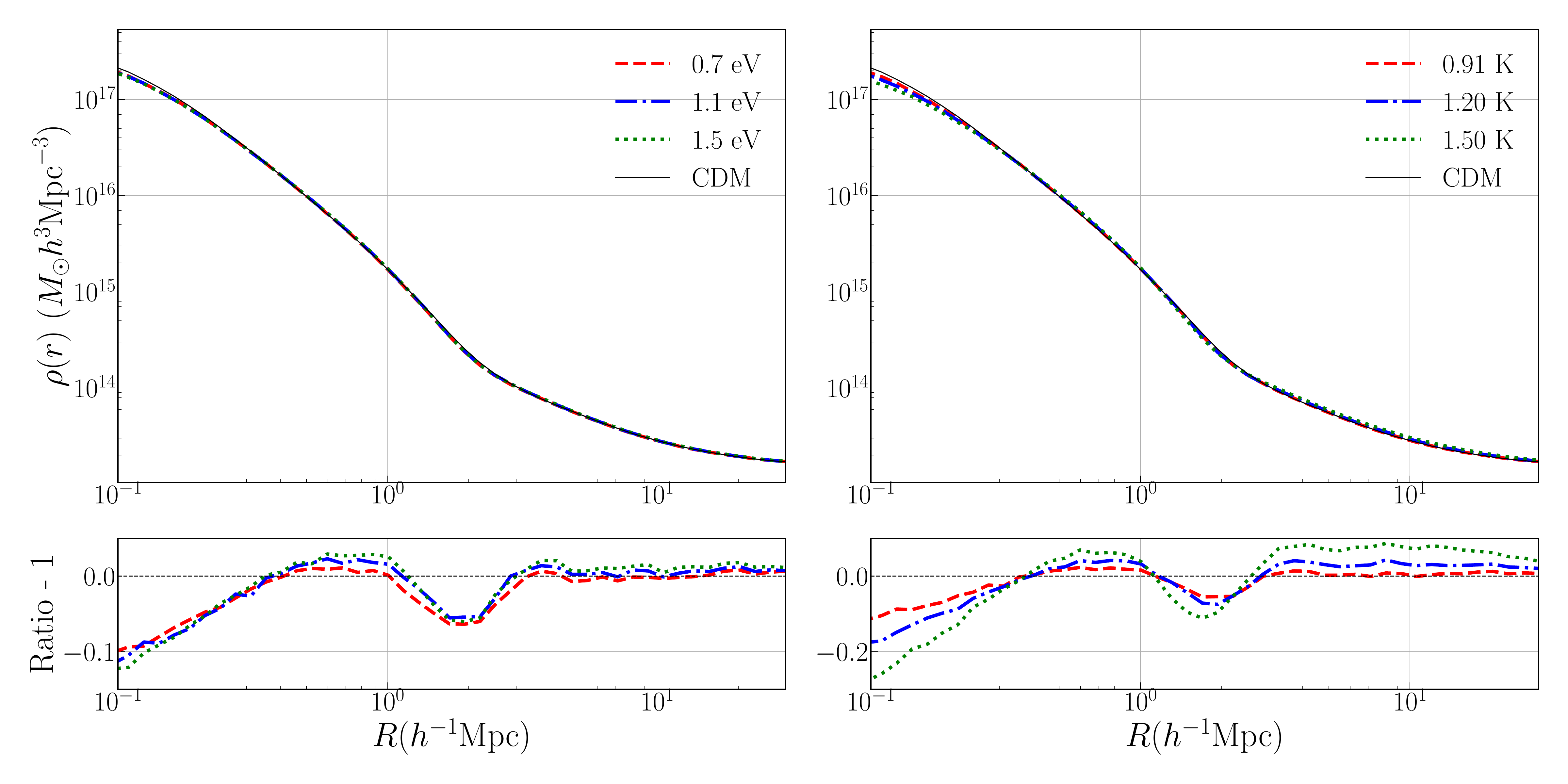}  
  \caption{Comparison of 3D density profiles for the CDM-only (1-fluid) case and various 3-fluid models. In the left column, the red dashed, blue dash–dotted, and green dotted lines correspond to LiMR temperatures fixed at 0.91 K with LiMR masses of 0.7 eV, 1.1 eV, and 1.5 eV, respectively. In the right column, the LiMR mass is fixed at 1.1 eV, while the temperatures are varied: 0.91 K (red dashed), 1.20 K (blue dash–dotted), and 1.50 K (green dotted). The total 3D profiles are shown in the 3-fluid case. The bottom panels present the relative comparison with the CDM-only (1-fluid) case.}
  \label{fig:3d_f}
\end{figure*}

\begin{figure*}
  \centering
  \includegraphics[width=\linewidth]{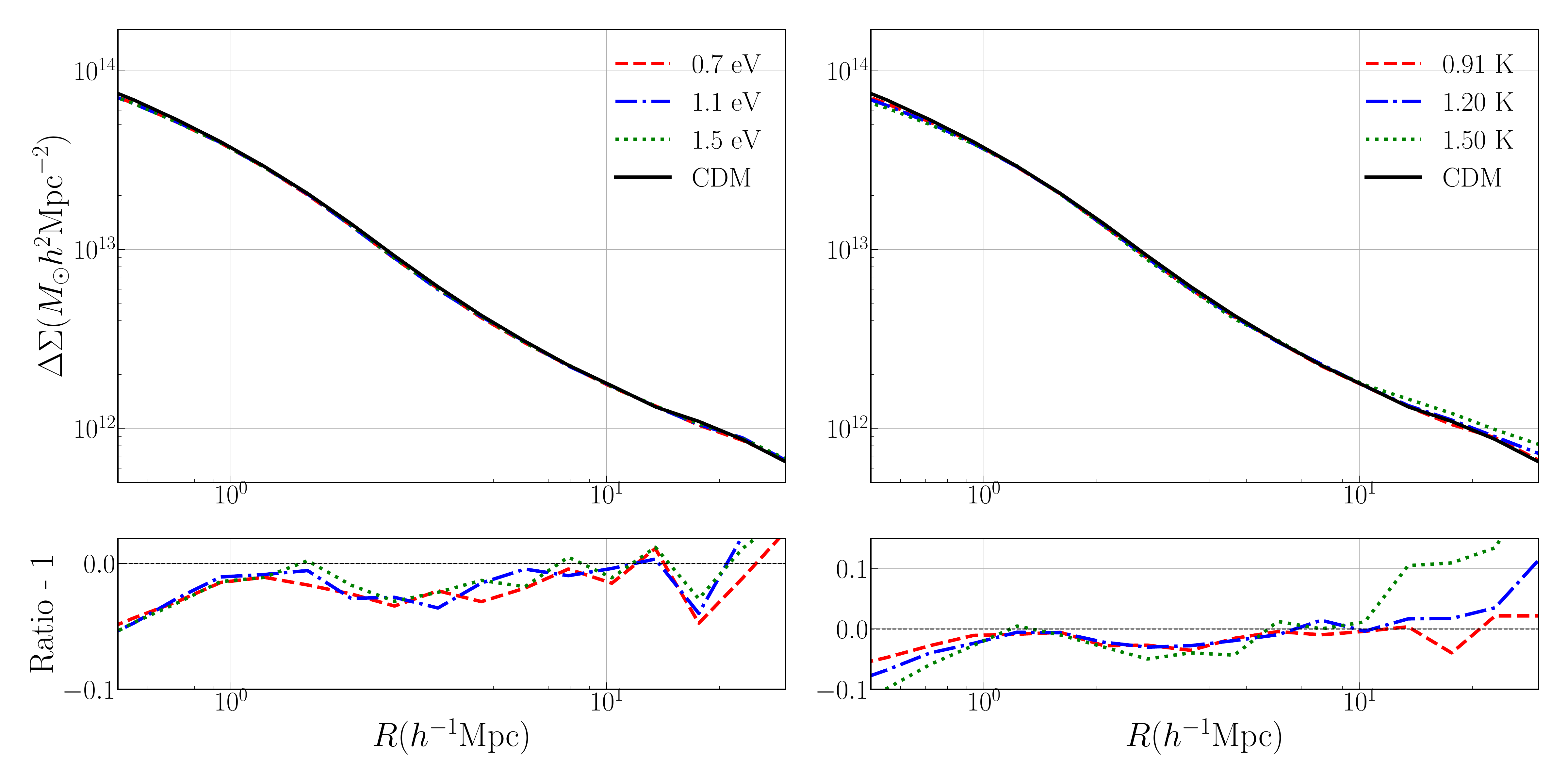}  
  \caption{Comparison of weak lensing in the CDM-only (1-fluid) case and various 3-fluid models. In the left column, the red dashed, blue dash–dotted, and green dotted lines represent models with LiMR temperature fixed at 0.91 K and LiMR masses of 0.7 eV, 1.1 eV, and 1.5 eV, respectively. In the right column, the LiMR mass is fixed at 1.1 eV, while the temperatures are varied: 0.91 K (red dashed), 1.20 K (blue dash–dotted), and 1.50 K (green dotted). The total weak lensing signal has been computed by including contributions from all three particle species in the 3-fluid case. The bottom panels show the relative comparison with the CDM-only (1-fluid) model.}
  \label{fig:lensing_f}
\end{figure*}

\subsection{Total Matter Power Spectrum}
We present the total matter power spectrum for 1-fluid, 2-fluid, and 3-fluid models. The total matter power spectrum is calculated as:
\begin{equation}
\begin{aligned}
    &  P_m(k) = \langle \delta_m(\vec{k}) \delta_m^*(\vec{k}) \rangle \quad \\
    \text{where,} \quad & \delta_m(\vec{k})=f_{cb} \delta_{cb}(\vec{k})+f_\nu \delta_\nu(\vec{k})+f_l \delta_l(\vec{k})
\end{aligned}    
\end{equation}
Here, $f_{i}$ and $\delta_{i}$ represent the fractional contribution and matter overdensities of $i^{th}$ species. Figure\,\ref{fig:pk_f} shows the comparison of the total matter power spectrum, $P_m(k)$, for the 1-fluid, 2-fluid, and 3-fluid cases. We first examine the impact of varying the mass of the LiMR particle (considered here as a Dirac fermion) on $P_m(k)$. The top-left panel of Figure\,\ref{fig:pk_f} displays the total matter power spectrum for five cases: (i) $\Omega_m = \Omega_{cb}$ in black (solid), (ii) $\Omega_m = \Omega_{cb} + \Omega_{\nu}$ in cyan (long-dashed), (iii) $\Omega_m = \Omega_{cb} + \Omega_{\nu} + \Omega_l(\rm M_l = 0.7,\mathrm{eV})$ in red (dashed), (iv) $\Omega_m = \Omega_{cb} + \Omega_{\nu} + \Omega_l(\rm M_l = 1.1,\mathrm{eV})$ in blue (dash-dotted), and (v) $\Omega_m = \Omega_{cb} + \Omega_{\nu} + \Omega_l(\rm M_l = 1.5,\mathrm{eV})$ in green (dotted). From Figure\,\ref{fig:pk_f}, it is evident that adding LiMR particles suppresses the total matter power spectrum, with the suppression becoming stronger as the LiMR mass increases. To highlight this effect, we plot the residuals, defined using the ratio of the power spectrum in 3-fluid cosmologies (CDM + baryons + SM neutrinos + LiMRs) relative to the 2-fluid case (CDM + baryons + SM neutrinos) as follows:-
\begin{equation}
    \text{Residual} \equiv \text{(Ratio - 1)} = \frac{P_m^{\rm {3-fluid}}(k)}{P_m^{\rm {2-fluid}}(k)} - 1
\end{equation}
As discussed earlier, this suppression arises from two competing effects: the effect of reducing CDM fraction and the effect of LiMR's clustering. The dominant contribution comes from the reduced CDM fraction, which drives the net suppression of $P_m(k)$. To clarify this point further, we also plot the LiMR auto-power spectrum, $P_l(k)$, in Figure\,\ref{fig:pk_f}. It can be seen that as the LiMR mass increases, $P_l(k)$ also increases, reflecting the fact that heavier particles cluster more efficiently. Nevertheless, the suppression in the total matter power spectrum is dominated by the reduction in the CDM component, which outweighs the enhanced clustering of the LiMRs.

In the top-right panel of Figure\,\ref{fig:pk_f}, we vary the temperature of the LiMR particles, keeping their mass fixed at 1.1 eV. We again plot the total matter power spectrum for five cases:  (i) $\Omega_m = \Omega_{cb}$ in black (solid), (ii) $\Omega_m = \Omega_{cb} + \Omega_{\nu}$ in cyan (long-dashed), (iii) $\Omega_m = \Omega_{cb} + \Omega_{\nu} + \Omega_l(\rm T_{l,0} = 0.91\,\mathrm{K})$ in red (dashed), (iv) $\Omega_m = \Omega_{cb} + \Omega_{\nu} + \Omega_l(\rm T_{l,0} = 1.20\,\mathrm{K})$ in blue (dash-dotted), and (v) $\Omega_m = \Omega_{cb} + \Omega_{\nu} + \Omega_l(\rm T_{l,0} = 1.50\,\mathrm{K})$ in green (dotted). As indicated by equation\,\eqref{eq:thermal_vel_f}, increasing the LiMR temperature enhances their thermal velocities, which leads to a stronger suppression in $P_l(k)$. To illustrate this, the right panel of Figure\,\ref{fig:pk_f} also shows the auto–power spectrum of the LiMR component. In this case, the effect of reducing CDM fraction and the effect of LiMR's clustering act in a complementary manner, producing an even larger suppression in the total matter power spectrum. Overall, Figure\,\ref{fig:pk_f} demonstrates that the suppression of the power spectrum is more sensitive to variations in the LiMR temperature than to changes in their mass. This arises because, when varying the mass, the competing effects of reduced CDM fraction and enhanced LiMR clustering partially cancel each other, whereas when varying the temperature, the two effects act in the same direction and reinforce one another.

\subsection{Halo Mass Funtion(HMF)}
The number density of DM halos as a function of their virial mass is known as the Halo-Mass function(HMF).
\begin{equation}
  \text { HMF } \equiv \frac{d n}{d \operatorname{log}_{10}\rm M} \quad ,  
\end{equation}
where $n$ denotes the halo number density and M the halo virial mass. The HMF encodes key information about structure formation in the Universe across different mass scales\,\cite{Tinker_2008} and is sensitive to several cosmological parameters, such as $\sigma_8$, $n_s$, and $\Omega_m$. We calculate HMF from the N-body simulations, using \texttt{Rockstar}\,\cite{Behroozi_2012} to generate halo catalogs from the \texttt{Gadget-3} snapshots at $z = 0$. In each simulation, we impose a threshold such that \texttt{Rockstar} identifies a halo only if it contains at least 40 particles. This allows us to define the minimum halo mass that can be reliably used in constructing the HMF. In our case, this minimum mass is set to $2.5 \times 10^{13} \, M_{\odot}$, corresponding to roughly 40 times the mean particle mass in simulations with LiMR particle masses of 0.7 eV, 1.1 eV, and 1.6 eV. The halo mass range is then divided into 20 logarithmically spaced bins between the minimum and maximum virial halo masses identified by \texttt{Rockstar}. We count the halos in each bin, normalize by the logarithmic bin width and the simulation volume (a cube of $1000 \, h^{-1}\,\mathrm{Mpc}$ on a side), and plot the resulting HMF against the bin centers. The shaded regions in the plots represent the $1\sigma$ uncertainty.

Figure\,\ref{fig:hmf_f} presents the HMFs from our simulations. In the left panel, we show results for LiMR particle masses of 0.7 eV in red (dashed), 1.1 eV in blue (dash-dotted), and 1.5 eV in green (dotted), all at a fixed temperature of $0.91 \, \mathrm{K}$ (the minimum-temperature scenario). These are compared with the 1-fluid case, keeping the total $\Omega_m$ constant across simulations. Note that all simulations with LiMRs are treated as 3-fluid models, including three species of Standard Model neutrinos and one cold dark matter (CDM) component. In the right panel, we fix the LiMR particle mass to 1.1 eV and vary the temperature to 0.91 K in red (dashed), 1.20 K in blue (dash-dotted), and 1.50 K in green (dotted). It is important to note that \texttt{Rockstar} catalogs only include halos composed of CDM particles, as the code is not sensitive to the LiMR or SM neutrino species present in the \texttt{Gadget-3} snapshots. Consequently, the derived HMFs do not account for the mass contributions from LiMRs and SM neutrinos, making the results indicative rather than fully accurate(this accounts for the effect of reduced CDM fraction only). Nonetheless, the trends are informative: we observe increasing suppression of the HMF as the LiMR particle mass increases. This suppression arises because higher LiMR masses correspond to larger values of $\Omega_l$, which in turn reduce the contribution from $\Omega_{cb}$.

\subsection{Mass-Concentration relation}

For a dark matter halo, the concentration is defined as:
\begin{equation}
    c=\frac{r_{\rm vir}}{r_s} \quad ,
\end{equation}
where, $r_{\rm vir}$ is the virial radius of the halo and $r_s$ is the scale radius. The virial radius is conventionally defined as the radius within which the mean density of the halo is 200 times the critical density of the Universe, while the scale radius corresponds to the point where the logarithmic slope of the density profile equals $-2$. The concentration thus characterizes the degree to which mass is centrally distributed within a halo\,\cite{okoli2017darkmatterhaloconcentrations, Correa_2015}. To study the mass–concentration relation, we divide the halo population into 10 logarithmically spaced bins between the minimum and maximum virial halo masses ($M_{\rm vir}$) identified by \texttt{Rockstar}. For each halo within a bin, we compute its concentration from the ratio $r_{\rm vir}/r_s$. The mean concentration of halos in each bin is then assigned to that mass bin. Figure\,\ref{fig:conc_f} presents the resulting relation between halo concentration and halo mass for different simulations. As expected, we find that lower-mass halos exhibit higher concentrations, while more massive halos are less concentrated. This behavior reflects the hierarchical growth of structure: smaller halos collapse earlier, when the cosmic background density was higher, resulting in denser central profiles. In contrast, massive halos typically assemble later, when the Universe is more diffuse, leading to lower concentrations. In Figure\,\ref{fig:conc_f} (top left panel), we observe that increasing the LiMR particle mass systematically reduces the concentration values across all mass bins. This trend is consistent with cosmological expectations: a larger LiMR mass increases $\Omega_l$, thereby lowering $\Omega_{cb}$, which suppresses clustering and produces less dense halos. A similar suppression is seen when varying the LiMR temperature. Since $\Omega_l \propto T_l^3$, higher temperatures yield larger $\Omega_l$, further reducing halo concentrations. This effect is evident in the top right panel of Figure\,\ref{fig:conc_f}.

\subsection{3D Density profiles}
The 3D mass density profile of a DM halo characterizes how mass is distributed as a function of radial distance from the halo center \,\cite{1996ApJ...462..563N, Navarro_1997, Moore_1998}. It is formally defined as:-
\begin{equation}
  \rho(r) \equiv \frac{1}{4\pi r^2} \frac{dM}{dr} \quad ,  
\end{equation}
where $M(r)$ is the total mass enclosed within a radius $r$.
To compute the 3D density profiles in our simulations, we use the \texttt{Rockstar} halo catalogs and select halos with virial masses between $10^{14}\, M_{\odot}$ and $2 \times 10^{14}\, M_{\odot}$. For each halo, we divide the radial distance from its center into 50 logarithmically spaced bins between $0.05 \, h^{-1}\,\mathrm{Mpc}$ and $30 \, h^{-1}\,\mathrm{Mpc}$. The mass contained in each radial bin is then divided by the shell volume,

\begin{equation}
  V_{\text{shell}} = \frac{4\pi}{3}\left(r_{i+1}^3 - r_i^3\right)  \quad ,  
\end{equation}
to estimate the density profile. This enables us to investigate how the relative structures of halos of comparable masses vary across different cosmologies.

In the 1-fluid case, this procedure directly yields the correct density profile since all mass is in the CDM component tracked by \texttt{Rockstar}. In the 3-fluid case, however, both LiMRs and SM neutrinos also contribute to the density distribution, but these species are not identified by \texttt{Rockstar}. To account for this, we use the same halo centers but compute the contribution from all three particle species. The total density profile is then obtained as

\begin{equation}
    \rho_{\text{total}}(r) = \rho_{cb}(r) + \rho_{\nu}(r) + \rho_{l}(r) \quad ,
\end{equation}
where $\rho_{cb}(r), \ \rho_{\nu}(r),$ and $\rho_{l}(r)$ represent the individual densities of CDM, SM neutrinos, and LiMR particles in different radial bins. Figure\,\ref{fig:3d_f} shows the total 3D density profiles for different cosmologies. Within a radius of $0.5 \, h^{-1}\,\mathrm{Mpc}$ from the halo centers, the density profiles in the 3-fluid runs are suppressed relative to the CDM-only case. The suppression grows stronger as the LiMR mass increases, corresponding to a lower effective $\Omega_c$. This trend is expected: LiMRs do not cluster as efficiently as cold dark matter, leading to reduced central halo densities. Moreover, the right panel of Figure\,\ref{fig:3d_f} demonstrates that increasing the LiMR temperature further enhances the suppression, consistent with the fact that hotter LiMRs free-stream more efficiently and therefore resist gravitational collapse. However, the overall trends are somewhat obscured because varying the cosmology also alters the hierarchy of structure formation. Consequently, comparing halos of the same mass across cosmologies may not be the most reliable approach. A more systematic analysis is required, which we leave for future work.

\begin{figure*}
  \centering
  \includegraphics[width=\linewidth]{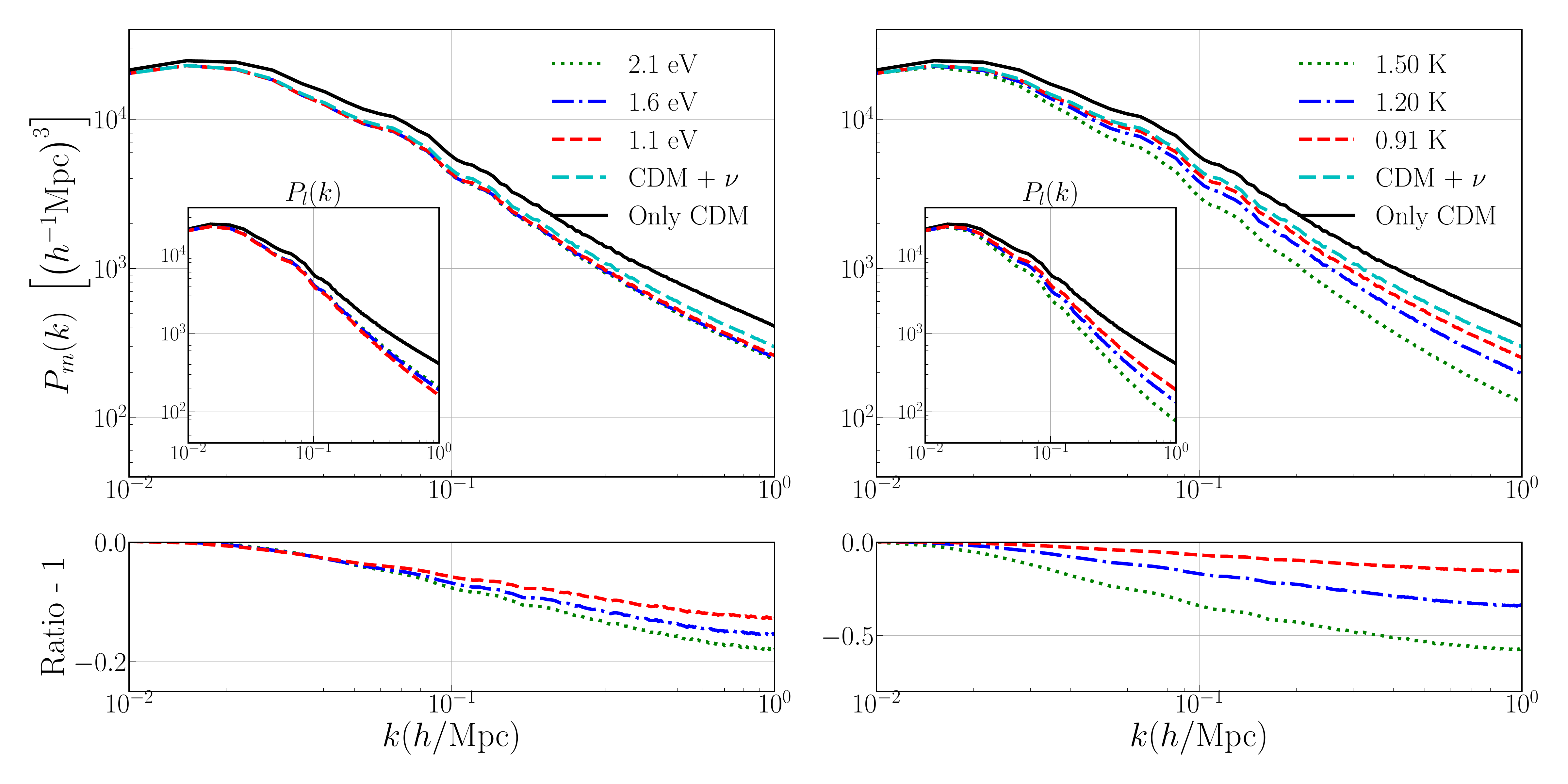}  
  \caption{In this plot, we examine the LiMR particle as a vector boson. The top-left panel shows the total matter power spectrum for five different cases: the black(solid) curve corresponds to a 1-fluid model (CDM only), the cyan(long dashed) curve to a 2-fluid model (CDM + SM neutrinos), and the red(dashed), blue(dash-dotted), and green(dotted) curves to 3-fluid models (CDM + SM neutrinos + LiMRs) with LiMR masses of 1.1 eV, 1.6 eV, and 2.1 eV, respectively. The inset displays the LiMR power spectrum for the three LiMR masses, with the black curve showing the CDM power spectrum in the 1-fluid case for comparison. The bottom-left panel compares the 3-fluid LiMR cosmologies with the 2-fluid case. On the right, the top panel fixes the LiMR mass at 1.6 eV and varies the LiMR temperature (0.91 K in red (dashed), 1.20 K in blue (dash-dotted), and 1.50 K in green (dotted)), while the bottom panel shows the residuals relative to the 2-fluid case. The corresponding inset highlights the LiMR power spectrum for the three different temperatures. Overall, these results demonstrate that the suppression in the power spectrum becomes stronger with increasing LiMR mass or temperature.}
  \label{fig:pk_b}

  \centering
  \includegraphics[width=\linewidth]{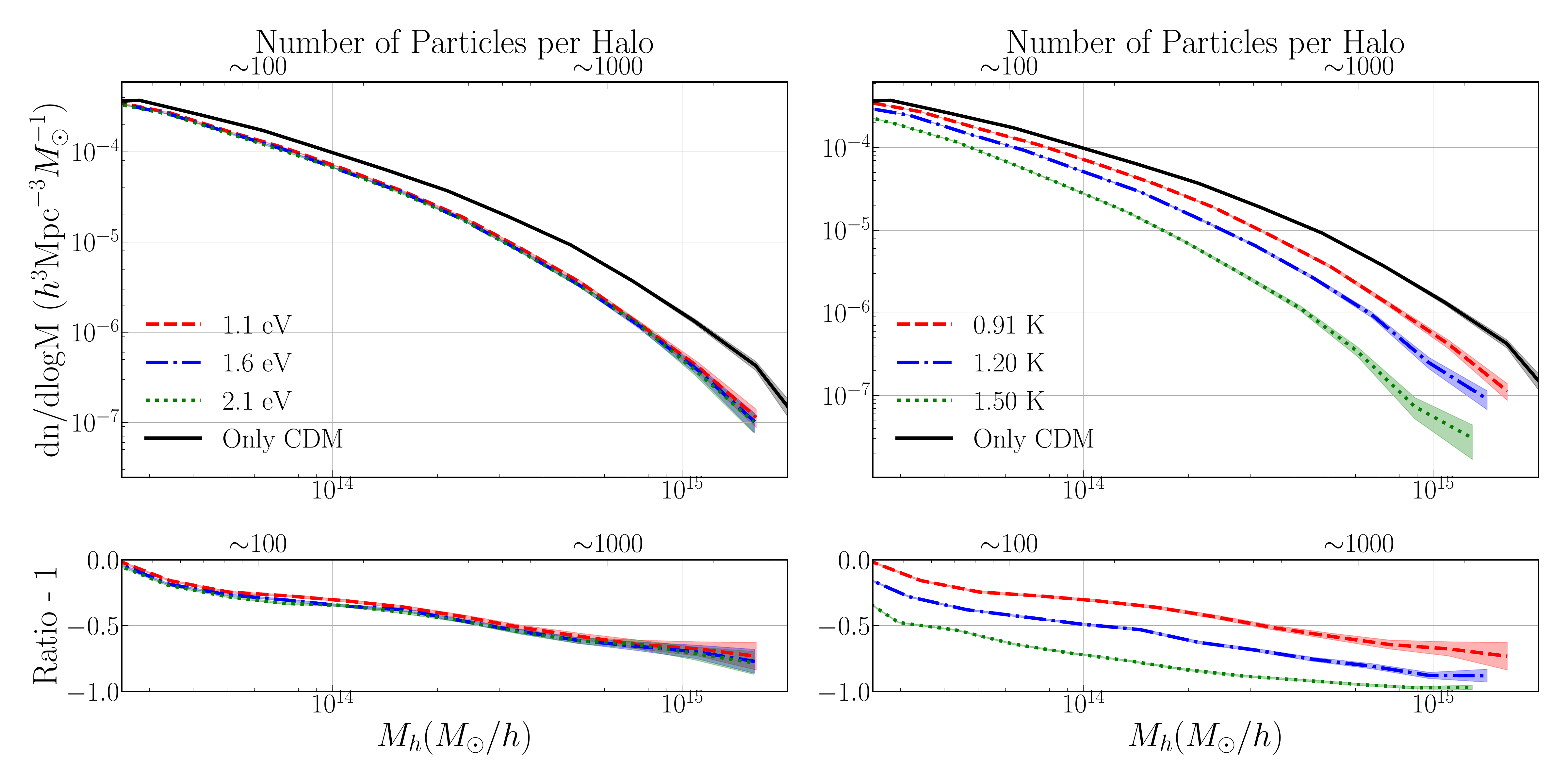}  
  \caption{Halo mass function for the 3-fluid model. The black solid line corresponds to a 1-fluid cosmology (CDM only, without SM neutrinos or LiMRs) and is compared to the 3-fluid case (CDM + SM neutrinos + LiMRs). In the left column, the red dashed, blue dash–dotted, and green dotted lines represent LiMR masses of 1.1 eV, 1.6 eV, and 2.1 eV, respectively (assuming vector boson). In the right column, the LiMR mass is fixed at 1.6 eV, while the temperature is varied: 0.91 K (red dashed), 1.20 K (blue dash–dotted), and 1.50 K (green dotted). The shaded regions indicate the $1\sigma$ uncertainties. The bottom row shows the ratio of the 3-fluid halo mass function to that of the 1-fluid case.}
  \label{fig:hmf_b}
\end{figure*}

\begin{figure*}
  \centering
  \includegraphics[width=\linewidth]{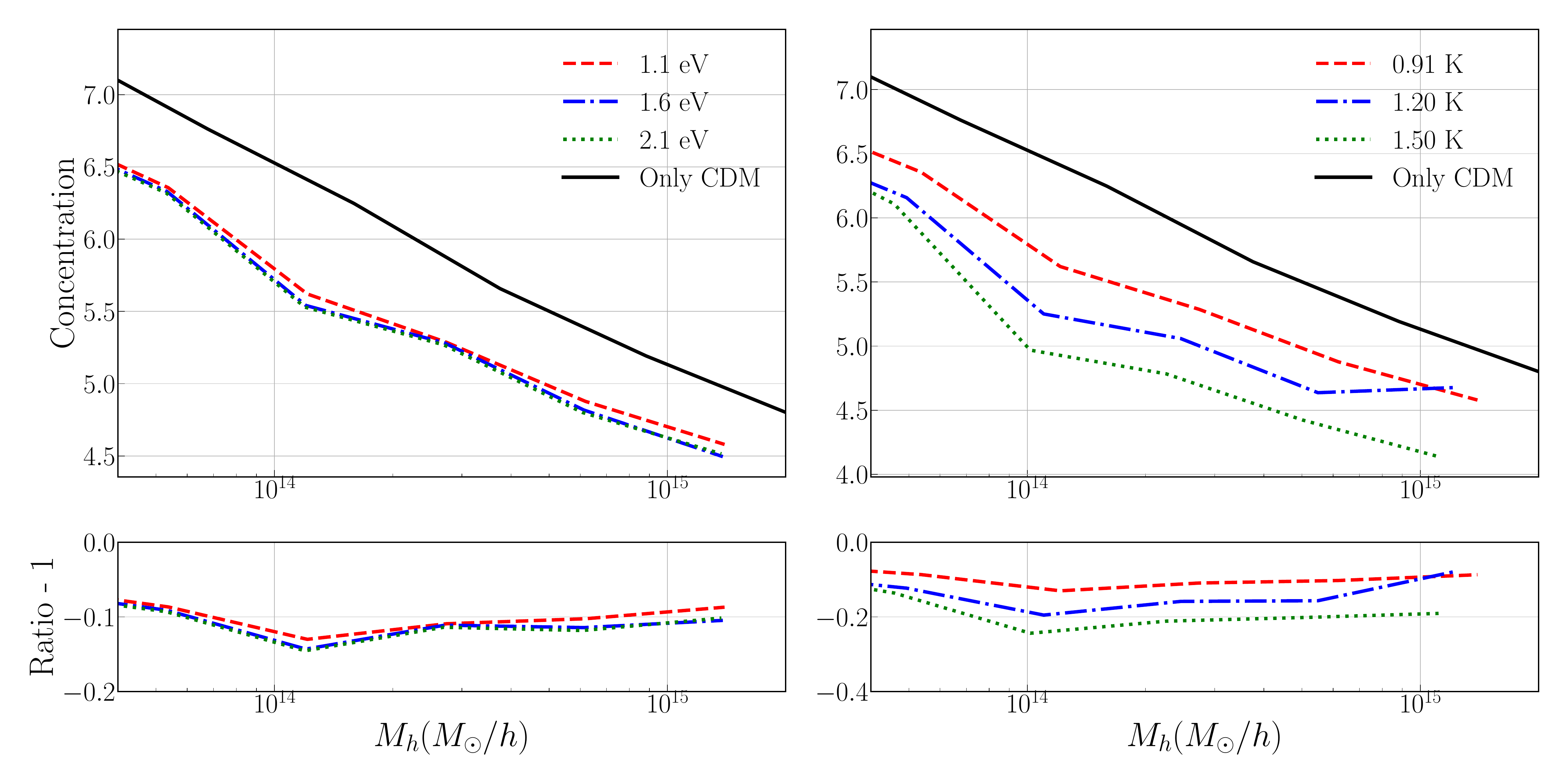}  
  \caption{Mass–concentration relation for the 3-fluid model. The black solid line corresponds to a cosmology with only CDM. In the top-left panel, the red dashed, blue dash-dotted, and green dotted lines represent the 3-fluid case (CDM + SM neutrinos + LiMRs) with a fixed LiMR temperature of 0.91 K and LiMR(vector boson in this case) masses of 1.1 eV, 1.6 eV, and 2.1 eV, respectively. The bottom-left panel shows the relative suppression of the mass–concentration relation in the 3-fluid model with respect to the CDM-only case. The results indicate that the suppression becomes stronger with increasing LiMR mass. In the top-right panel, the red dashed, blue dash-dotted, and green dotted lines correspond to the 3-fluid model with a fixed LiMR mass of 1.6 eV and temperatures of 0.91 K, 1.20 K, and 1.50 K, respectively. The bottom-right panel shows the corresponding relative suppression with respect to the CDM-only (1-fluid) case.}
  \label{fig:conc_b}

  \centering
  \includegraphics[width=\linewidth]{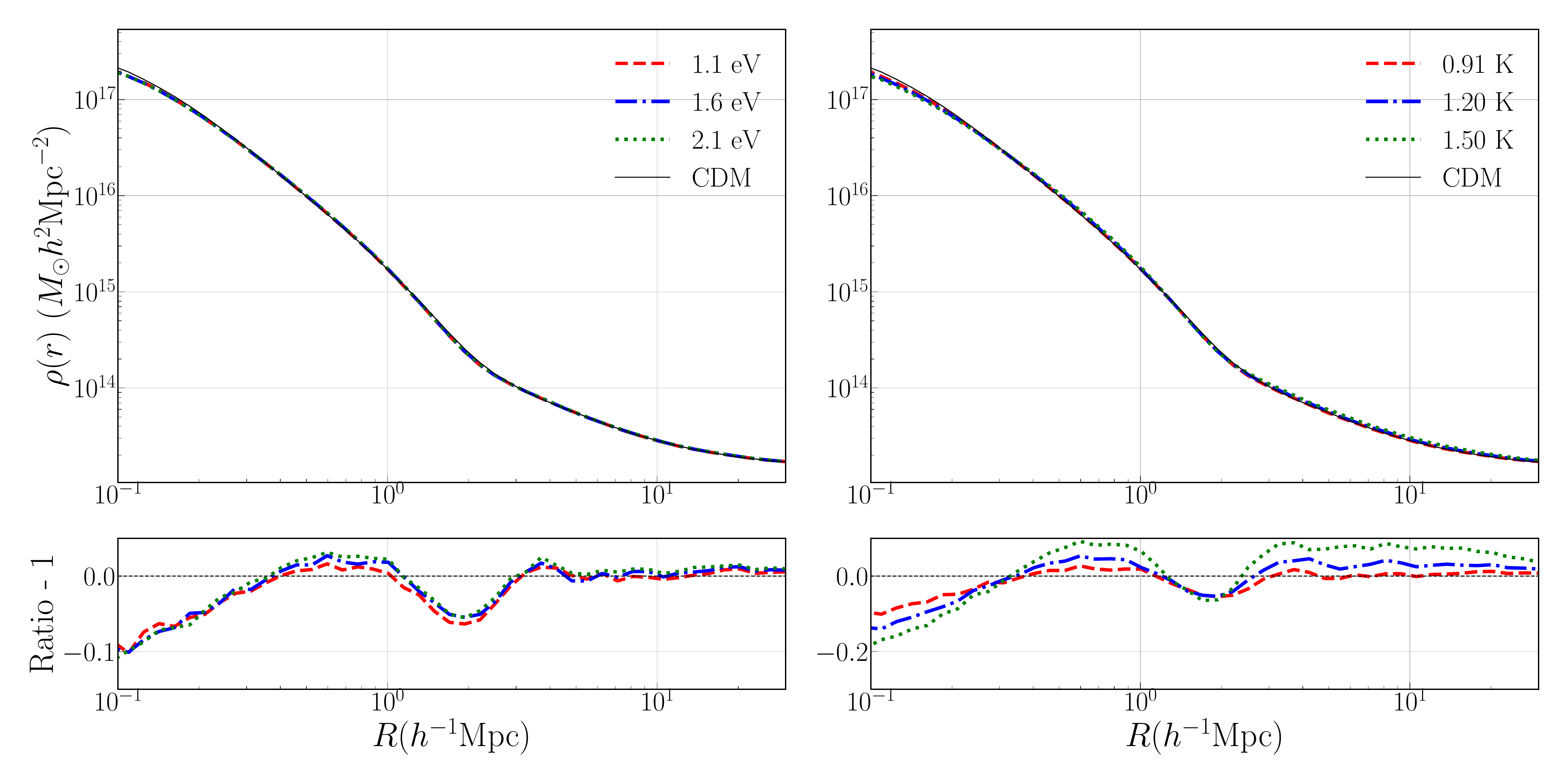}  
  \caption{Comparison of 3D density profiles(for vector bosons) for the CDM-only (1-fluid) case and various 3-fluid models. In the left column, the red dashed, blue dash–dotted, and green dotted lines correspond to LiMR temperatures fixed at 0.91 K with LiMR masses of 1.1 eV, 1.6 eV, and 2.1 eV, respectively. In the right column, the LiMR mass is fixed at 1.6 eV, while the temperatures are varied: 0.91 K (red dashed), 1.20 K (blue dash–dotted), and 1.50 K (green dotted). The total 3D profiles are shown in the 3-fluid case. The bottom panels present the relative comparison with the CDM-only (1-fluid) case.}
  \label{fig:3d_b}
\end{figure*}

\begin{figure*}
  \includegraphics[width=\linewidth]{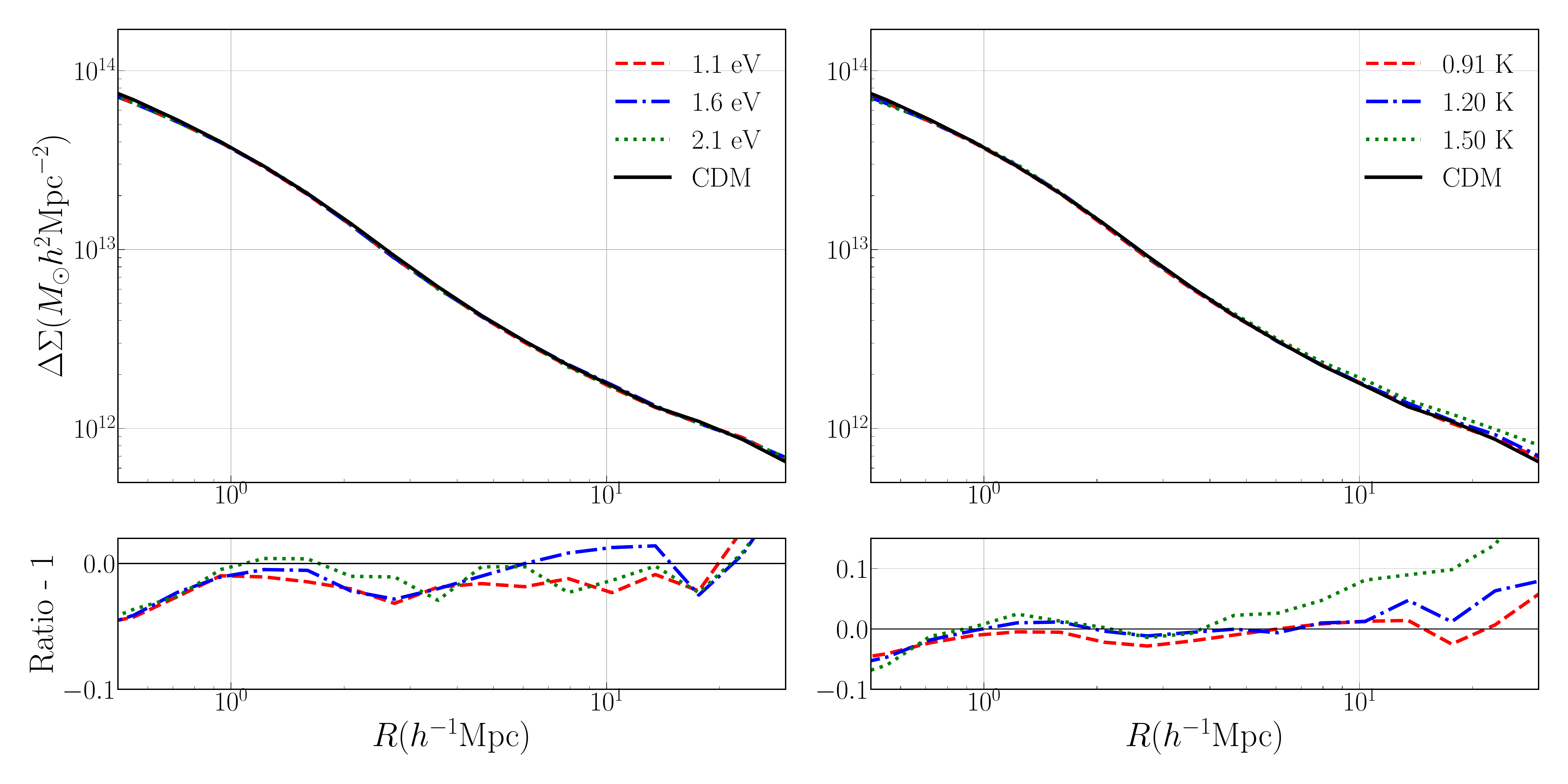}  
  \caption{Comparison of weak lensing(for vector bosons) in the CDM-only (1-fluid) case and various 3-fluid models. In the left column, the red dashed, blue dash–dotted, and green dotted lines represent models with LiMR temperature fixed at 0.91 K and LiMR masses of 1.1 eV, 1.6 eV, and 2.1 eV, respectively. In the right column, the LiMR mass is fixed at 1.6 eV, while the temperatures are varied: 0.91 K (red dashed), 1.20 K (blue dash–dotted), and 1.50 K (green dotted). The total weak lensing signal has been computed by including contributions from all three particle species in the 3-fluid case. The bottom panels show the relative comparison with the CDM-only (1-fluid) model.}
  \label{fig:lensing_b}
\end{figure*}

\subsection{Weak lensing}
In weak lensing surveys, the key observable derived from galaxy images is the tangential shear\,\cite{Troxel_2018}. This quantity can be directly related to the excess surface mass density($\Delta \Sigma$) around a dark matter halo or cluster. Thus, $\Delta \Sigma$ serves as a bridge between observational data and theoretical predictions, making it one of the most widely used methods to map the matter distribution around massive halos or clusters\,\cite{Munshi_2005}.

The excess surface mass density in simulations can be computed as:
\begin{equation}
    \Delta \Sigma(R) = \Sigma(<R) - \Sigma(R) \quad ,
\end{equation}
where R is the projected 2D distance from the halo (or cluster) center. Here, $\Sigma(<R)$ denotes the mean surface mass density within radius R, and $\Sigma(R)$ corresponds to the local surface mass density at radius R.
To calculate $\Delta \Sigma(R)$ in simulations, we use halos identified in the \texttt{Rockstar} catalogs with masses between $10^{14}  M_\odot$ and $2 \times 10^{14} M_\odot$. For each halo, we count the mass enclosed within a cylinder of height $70 \ h^{-1}\mathrm{Mpc}$, aligned along the line-of-sight (LoS) axis and centered on the halo.

\begin{enumerate}
    \item The first term, $\Sigma(<R)$, is obtained by summing the mass of particles enclosed within radius R and dividing by the projected circular area $\pi R^2$.
    \item The second term, $\Sigma(R)$, is computed by considering a thin annulus of width $dr = 0.01 \ h^{-1}\mathrm{Mpc}$ around R. The enclosed mass in this annulus is divided by its projected surface area, $2 \pi R dr$.
\end{enumerate}
We perform this calculation in 25 logarithmically spaced radial bins between $0.05 \ h^{-1}\mathrm{Mpc}$ and $30 \ h^{-1}\mathrm{Mpc}$. The resulting $\Delta \Sigma(R)$ is then averaged (stacked) over all halos in the mass range $10^{14} \ M_\odot$ to $2 \times 10^{14}  M_\odot$ for each simulation box. It is important to note that \texttt{Rockstar} identifies halos using only Type-1 particles (CDM). Therefore, the halo mass it reports includes contributions from CDM alone. However, in simulations that also include Standard Model (SM) neutrinos and LiMRs, we compute the enclosed mass using the contributions from all three particle species. In this case, the total excess surface mass density is given by:
\begin{equation}
    \Delta \Sigma_{\text{total}} = \Delta \Sigma_{cb} + \Delta \Sigma_\nu + \Delta \Sigma_l \quad .
\end{equation}
This $\Delta \Sigma_{\text{total}}$ is the quantity plotted against logarithmic radial bins in Figure\,\ref{fig:lensing_f}. Like the 3D density profiles of halos, the trends are somewhat obscured and require further analysis.

\subsection{A Bosonic species as LiMR}
In the previous sections, we considered the case where the LiMR particle is a fermion. Here, we also explore the possibility that LiMRs are bosons. As discussed in Section II, the free-streaming scale, matter density parameter, and thermal velocities all depend on whether the LiMR is a fermionic or bosonic particle. For the bosonic case, we model LiMRs as vector bosons and run a new set of five N-body simulations. In the first set, we fix the LiMR temperature to the minimum scenario of 0.91 K and vary the particle mass, taking values of 1.1 eV, 1.6 eV, and 2.1 eV. In the second set, we fix the LiMR mass at 1.6 eV and vary the temperature to 1.20 K and 1.50 K, following the same procedure as in the fermionic case.

We then examine the impact of these variations in mass and temperature on large-scale structure formation. We focus on five observables for bosonic LiMRs: the total matter power spectrum in Figure\,\ref{fig:pk_b}, the halo mass function in Figure\,\ref{fig:hmf_b}, the halo mass–concentration relation in Figure\,\ref{fig:conc_b}, 3D density profiles in Figure\,\ref{fig:3d_b}, and weak lensing signals in Figure\,\ref{fig:lensing_b}. For the bosonic case, we observe qualitatively similar trends: the total matter power spectrum shows suppression across all scales, with stronger suppression for higher masses and temperatures. The HMF and the mass–concentration relation also display suppression consistent with the fermionic case. For 3D density profiles and weak lensing, the suppression is less pronounced but follows the same qualitative trend. Beyond these similarities, it is particularly interesting to examine how changes in the underlying distribution function of LiMR particles affect cosmological observables. A detailed analysis of this effect will be addressed in future studies. Here, our aim is to demonstrate that the pipeline we have developed is capable of incorporating such complications in a consistent manner. Establishing this capability is crucial, as different distribution functions arise naturally in many particle-physics scenarios, and their imprints on cosmological observables may provide a pathway to distinguish between competing models of new physics.

\section{Conclusions}
We have presented the first results from the fully non-linear evolution of a mixed dark matter scenario that includes three particle species: Cold Dark Matter (CDM), Standard Model (SM) neutrinos, and Light Massive Relics (LiMRs). While CDM remains the dominant matter component in the Universe, the presence of LiMRs, in addition to SM neutrinos, introduces distinct signatures in large-scale structure formation. ``In particular, LiMRs with masses of a few eV occupy the most compelling region of parameter space, as they are neither heavy enough to cluster like CDM nor light enough to remain relativistic until the present epoch, thereby influencing structure formation in a uniquely intermediate manner." The impact of various LiMR properties—such as their free-streaming scales, distribution functions, and contributions to $\Delta N_{\text{eff}}$—on structure formation and cosmological observables has been explored in this work. Our results demonstrate that these properties can leave distinct and measurable imprints on nonlinear structures, motivating more detailed, high-resolution studies to fully quantify their cosmological signatures. Our main findings can be summarized as follows:

\begin{enumerate}[label=\arabic*)]
\item \textit{Effects on structure formation}:
The density field exhibits a noticeable washing-out of small-scale structures at late times in the presence of LiMRs. Importantly, while LiMRs cluster more efficiently than SM neutrinos, their clustering amplitude remains below that of CDM, in agreement with expectations from free-streaming arguments.
\item \textit{Matter power spectrum}: The total matter power spectrum shows enhanced suppression relative to the CDM+neutrino case. This suppression is more sensitive to the temperature of LiMRs than to their mass. This arises because the relic abundance of LiMRs scales as $T^3$, while depending only linearly on their mass. 
\item \textit{Halo statistics}: Using the \texttt{Rockstar} halo catalogs, we examined the halo mass function (HMF) and mass–concentration relations. Both quantities exhibit suppression when LiMRs are included. We emphasize that, as in most halo-finding methods, halo properties are defined only with respect to the CDM component, while LiMR and neutrino contributions are not directly included.
\item \textit{3D density profiles and weak lensing}: The density profiles, $\rho(r)$, as well as the excess surface mass density, $\Delta\Sigma$, exhibit a clear suppression in the 3-fluid scenario within a radius of $0.5 \, h^{-1}\,\mathrm{Mpc}$ from the halo (or cluster) center. Although a more detailed analysis is required, the results suggest that a non-negligible fraction of LiMR particles may virialize and reside in the vicinity of halo centers. This opens an interesting avenue for future investigation, as weak-lensing measurements could be sensitive to these subtle but cumulative effects.
\item \textit{Fermionic vs. Bosonic LiMRs}: We further explored the case of LiMRs as bosons (vector particles), in contrast to the fermionic case(Dirac fermion) considered initially. While the overall qualitative trends remain the same, subtle differences in clustering and suppression could arise. A detailed comparison between the fermionic and bosonic cases is left to future work.  

\end{enumerate}

In this first paper of a planned series, our primary aim has been to establish the simulation pipeline and demonstrate its ability to handle the complexities of multi-fluid cosmologies. Developing such a framework is important not only for assessing the detectability of LiMRs with upcoming cosmological surveys but also for breaking degeneracies with massive neutrinos and for providing new avenues to connect particle physics models of light relics with cosmological data. 

\section*{ACKNOWLEDGMENTS}
AB’s work was partially supported by the Startup Research Grant (SRG/2023/000378) from the Science and Engineering Research Board (SERB), India. VS acknowledges financial support from the University Grants Commission (UGC), India, through the UGC–SRF fellowship. The authors acknowledge the PARAMBrahma Facility under the National Supercomputing Mission, Government of India, at the Indian Institute of Science Education and Research, Pune, for providing the computing resources for this work. We also acknowledge the use of the Pylians\,\cite{2018asclsoft11008V} library for numerical analysis, and thank its developers for making it publicly available. This work also made extensive use of NumPy\,\cite{harris2020array}, Scipy\,\cite{virtanen2020scipy}, and Matplotlib\,\cite{hunter2007matplotlib} for data processing and visualization.

\newpage

\section*{Appendix}

\begin{figure*}
  \centering
  \includegraphics[width=\linewidth]{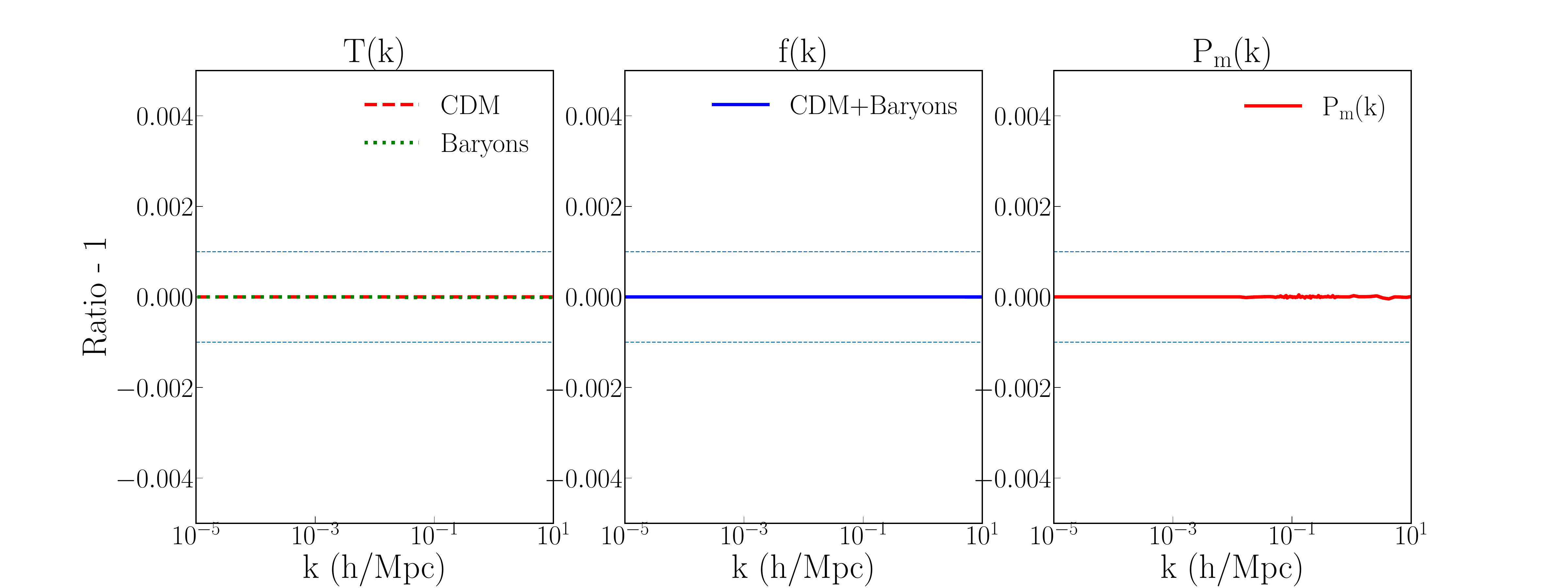}  
  \caption{In this plot, we compare the transfer functions, growth rates, and the total matter power spectrum between \texttt{Py-RePS} and \texttt{RePS} for the only CDM case(1-fluid). The residual(i.e., Ratio - 1) has been plotted for all three quantities. The results from the two codes match within the machine precision.}
  \label{fig:reps_1f}
  
  \centering
  \includegraphics[width=\linewidth]{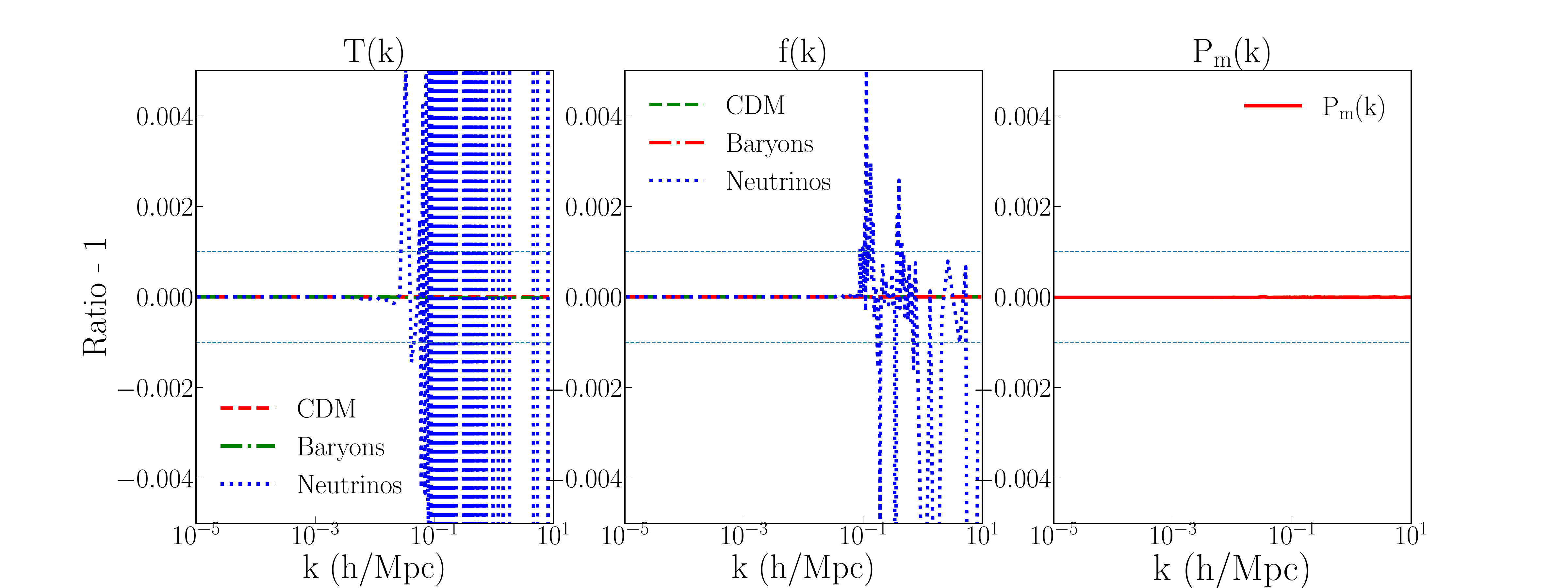}  
  \caption{In this plot, we compare the transfer functions, growth rates, and the total matter power spectrum between \texttt{Py-RePS} and \texttt{RePS} for the CDM + SM neutrinos case(2-fluid). The residual(i.e., Ratio - 1) has been plotted for all three quantities. The results from the two codes match within the machine precision.}
  \label{fig:reps_2f}
\end{figure*}

\begin{figure*}
  \centering
  \includegraphics[width=\linewidth]{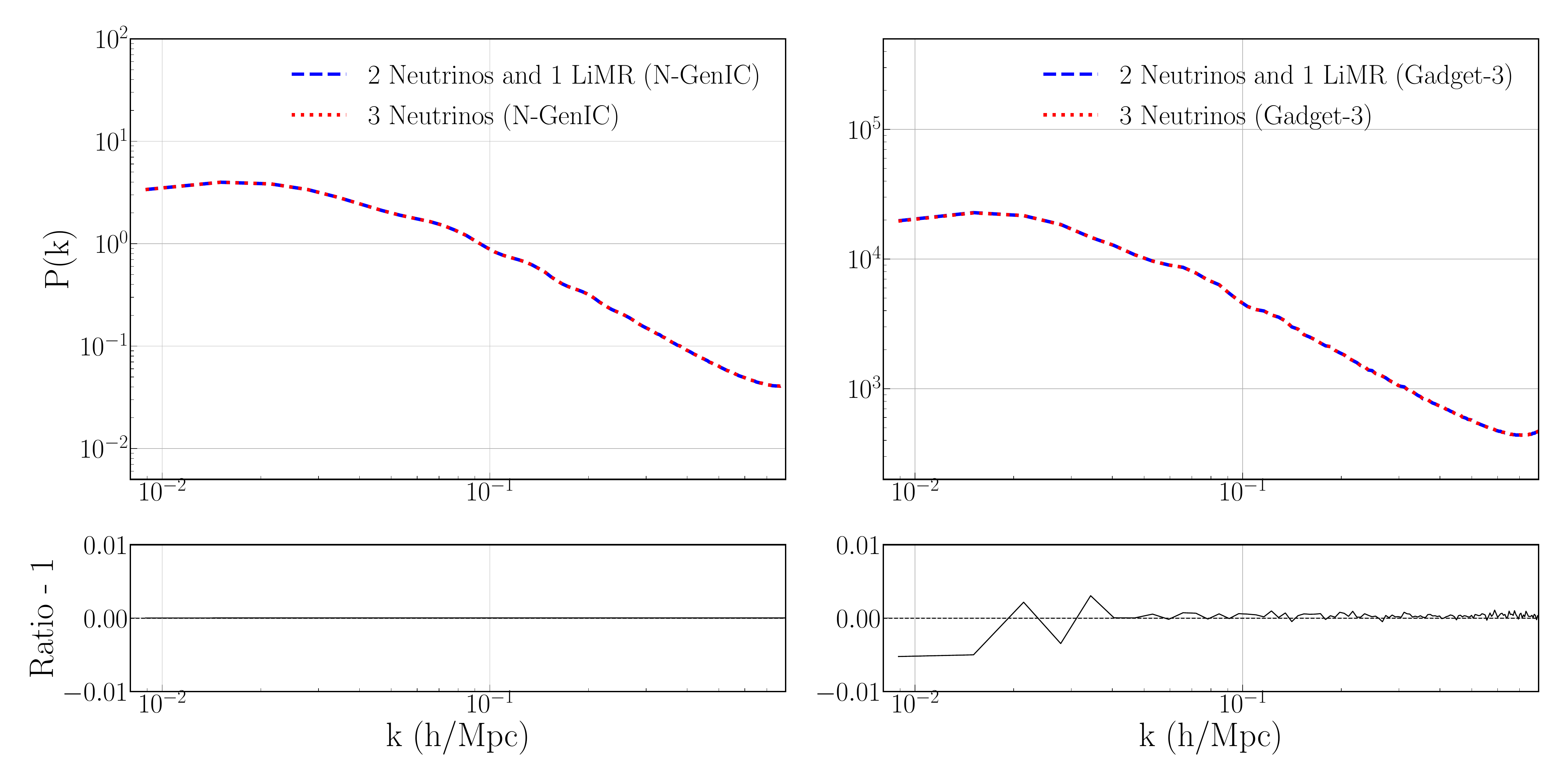}  
  \caption{\textit{Check for Modified} \texttt{N-GenIC} : Left plot shows the comparison of the total matter power spectrum at redshift 99 generated using \text{N-GenIC}. The red(dotted) line represents the $P_m(k)$ for the case when we have 3 neutrinos, and the blue(dashed) line is for the case when we pass two neutrinos with the neutrino flag and the third neutrino as a LiMR particle. \\ \textit{Check for Modified} \texttt{Gadget-3} : Right plot shows the comparison of $P_m(k)$ at redshift 0 generated using the \texttt{Gadget-3} code after running the simulations. The plot below represents the (Ratio - 1) for the two power spectras shown in the plot above them. This shows that the $P_m(k)$ match in these two cases(both at initial and final redshifts), and hence the 3-fluid pipeline works perfectly.}
  \label{fig:consistency_check}
\end{figure*}

\subsection{Matching RePS}
We have developed the model and formulated the necessary equations\,(\ref{eq:c})–(\ref{eq:d}) to perform the backscaling procedure, following the methodology implemented in \texttt{RePS}. To this end, we have written a Python-based counterpart of \texttt{RePS} (originally implemented in C), which we refer to as \texttt{Py-RePS}.

As a first validation step for our 3-fluid implementation, we benchmark the code against the existing 2-fluid version of \texttt{RePS}, testing both the 1-fluid and 2-fluid scenarios. Specifically, we compare the residuals of the matter power spectrum $(P(k))$, transfer functions $(T(k))$, and growth factors $({\rm f}(k))$, defined respectively as

$$
\frac{P(k)_{\rm Py-RePS}}{P(k)_{\rm RePS}} - 1, \quad  
\frac{T(k)_{\rm Py-RePS}}{T(k)_{\rm RePS}} - 1, \quad  
\frac{{\rm f}(k)_{\rm Py-RePS}}{{\rm f}(k)_{\rm RePS}} - 1.
$$

Figure\,\ref{fig:reps_1f} shows the comparison between \texttt{RePS} and \texttt{Py-RePS} for the 1-fluid case (CDM only). The two codes agree to better than $0.01\%$, for a cosmology as specified in section III.

We next validate the code for the 2-fluid case, in which we include, in addition to CDM, three neutrino species with masses of $0.1\,\mathrm{eV}$ each. The total matter density parameter $\Omega_m$ is kept fixed, with $\Omega_c = \Omega_m - \Omega_\nu$, and the neutrino contribution $\Omega_\nu$ calculated using Eq.\,\eqref{eq:omega_f}. Figure\,\ref{fig:reps_2f} presents the comparison of $P(k)$, $T(k)$, and ${\rm f}(k)$ between the two codes. In this case as well, \texttt{Py-RePS} reproduces the results of \texttt{RePS} to machine precision.

These tests validate the accuracy of \texttt{Py-RePS} against the state-of-the-art \texttt{RePS} code for initializing massive neutrino N-body simulations. Having established this consistency, we extend \texttt{Py-RePS} to the 3-fluid case using the set of equations described in Section II. \\

\subsection{Consistency check for 3-Fluid case}
After modifying all the codes required to run 3-fluid simulations, we perform a consistency check of the complete pipeline to ensure its correctness. We consider two scenarios. In the first, the N-body simulations include CDM and three standard model (SM) neutrino species of equal mass. In the second, we include CDM, two SM neutrinos, and treat the third neutrino as a LiMR particle. In both cases, the total energy density of CDM and neutrinos is kept fixed, and the physical properties of the third neutrino (mass, temperature, and velocities) are assigned to the LiMR particle. In principle, these two scenarios are physically equivalent; thus, this test provides a stringent validation of the modified 3-fluid simulation pipeline. If the modifications are implemented correctly, the two scenarios should yield identical results.

Figure\,\ref{fig:consistency_check} presents the comparison. The left panel shows the total matter power spectrum at redshift $z=99$, generated from the initial conditions produced by the modified \texttt{N-GenIC} code. The right panel shows the corresponding results from the \texttt{Gadget-3} simulations at redshift $z=0$. In both cases, the agreement between the two scenarios is at the sub-percent level. This confirms the consistency of the full pipeline and establishes the reliability of our 3-fluid N-body simulations.


\bibliographystyle{apsrev4-2}

\bibliography{main}

\end{document}